\documentclass[onecolumn]{IEEEtran}
\usepackage{amsmath,amssymb,amsthm,cite,multirow,url,graphicx}
\usepackage[algo2e]{algorithm2e}

\theoremstyle{definition}

\providecommand{\GF}{\mathrm{GF}}

\fussy

\title{Reduced-Complexity Reed--Solomon Decoders Based on Cyclotomic FFTs}
\author{
Ning~Chen,~\IEEEmembership{Student Member,~IEEE} and Zhiyuan~Yan,~\IEEEmembership{Senior Member,~IEEE}
\thanks{This work was supported in part by Thales Communications Inc. and in part by a grant from the Commonwealth of Pennsylvania, Department of Community and Economic Development, through the Pennsylvania Infrastructure Technology Alliance (PITA).
The material in this paper was presented in part at the IEEE
Workshop on Signal Processing Systems, Shanghai, China, October 2007.
}
\thanks{The authors are with the Department of ECE, Lehigh University,
Bethlehem, PA 18015 USA (E-mails: \{nic6, yan\}@lehigh.edu).} }
\begin{document}
\maketitle
 \begin{abstract}
In this paper, we reduce the computational complexities of partial
and dual partial cyclotomic FFTs (CFFTs), which are discrete Fourier
transforms where spectral and temporal components are constrained,
based on their properties as well as a common subexpression
elimination algorithm. Our partial CFFTs achieve smaller
computational complexities than previously proposed partial CFFTs.
Utilizing our CFFTs in both transform- and time-domain
Reed--Solomon decoders, we achieve significant complexity reductions.
\end{abstract}
\section{Introduction}
Due to the widespread applications of Reed--Solomon (RS) codes~\cite{Blahut83} in various digital communication and storage
systems, efficient RS decoding has been an important
research topic (see, for example,~\cite{Jeng99,Costa04,Truong06a,Lin07,Fedorenko06,Fedorenko02,Truong01}).
Since all syndrome-based hard-decision decoding methods for RS codes
involve discrete Fourier transforms (DFTs) over finite fields~\cite{Blahut83},
fast Fourier transform (FFT) algorithms can be used to reduce the complexity of RS
decoders (see, for example,~\cite{Costa04,Truong06a,Fedorenko06}).

Using an approach similar to those in previous works (see, for
example,~\cite{Zakharova92}), cyclotomic FFTs (CFFTs) were recently
proposed~\cite{Trifonov03} and two variants were subsequently
considered~\cite{Costa04,Fedorenko06}. To avoid confusion,
in this paper we refer to the CFFTs proposed in~\cite{Trifonov03} as direct CFFTs (DCFFTs) and those in~\cite{Costa04} and \cite{Fedorenko06} as inverse CFFTs
(ICFFTs) and symmetric CFFTs (SCFFTs), respectively. Given
a primitive element $\alpha \in \mathrm{GF}(2^m)$, the DFT of a
vector $\boldsymbol{f} = (f_0,f_1,\dots,f_{n-1})^T$ is defined as
$\boldsymbol{F}\triangleq\bigl(f(\alpha^0), f(\alpha^1), \dots,
f(\alpha^{n-1})\bigr)^T$, where $f(x) \triangleq \sum_{i=0}^{n-1}f_i x^i
\in \mathrm{GF}(2^m)[x]$.
A DCFFT is given by $\boldsymbol{F}=\boldsymbol{ALf'}
=\boldsymbol{AQ}(\boldsymbol{c}\cdot \boldsymbol{Pf}')$,
where $\boldsymbol{A}$ is an $n\times n$ binary matrix,
$\boldsymbol{L}$ is a block diagonal matrix, $\boldsymbol{f}'$ is a
permutation of the input vector $\boldsymbol{f}$,
$\boldsymbol{c}$ is a pre-computed vector, $\cdot$ stands for
pointwise multiplications, and $\boldsymbol{Q}$ and
$\boldsymbol{P}$ are both sparse binary matrices. Similarly, an
SCFFT is given by
$\boldsymbol{F}'=\boldsymbol{L}^T\boldsymbol{A}'^{T}
\boldsymbol{f}'=\boldsymbol{P}^T\bigl(\boldsymbol{c}\cdot(\boldsymbol{A}'\boldsymbol{Q})^T
\boldsymbol{f}'\bigr)$,
where both
$\boldsymbol{F}'$ and $\boldsymbol{f}'$ are permuted by the same
permutation matrix, and $\boldsymbol{L}^T\boldsymbol{A}'^{T}$ is
symmetric. Finally, based on inverse DFTs, an ICFFT is given by
$\boldsymbol{F}''=\boldsymbol{L}^{-1}\boldsymbol{A}^{-1}\boldsymbol{f}
=\boldsymbol{P}^T(\boldsymbol{c^*}\cdot
\boldsymbol{Q}^T\boldsymbol{A}^{-1}\boldsymbol{f})$,
where $\boldsymbol{F}''$ is also a permutation of $\boldsymbol{F}$
and $\boldsymbol{c^*}$ is a pre-computed vector.
Since all CFFTs are in bilinear forms~\cite{Blahut83}, we refer to
$\boldsymbol{P}$, $(\boldsymbol{A}'\boldsymbol{Q})^T$, and
$\boldsymbol{Q}^T\boldsymbol{A}^{-1}$ as \emph{pre-addition} matrices and
$\boldsymbol{AQ}$, $\boldsymbol{P}^T$, and $\boldsymbol{P}^T$ as
\emph{post-addition} matrices for DCFFTs, SCFFTs, and ICFFTs, respectively.
The numbers of non-one elements in $\boldsymbol{c}$ or
$\boldsymbol{c^*}$ are the number of multiplications required, and
the pre- and post-addition matrices determine the additive
complexities of CFFTs. Though CFFTs in~\cite{Trifonov03,Costa04,Fedorenko06} achieve low multiplicative
complexities, their additive complexities (numbers of additions
required) are very high, with or without the various methods used in~\cite{Trifonov03,Costa04,Fedorenko06} to reduce the additive
complexities. In our previous work~\cite{Chen08a}, we proposed a
novel common subexpression elimination (CSE) algorithm, and then
used it to reduce the additive complexities of \textbf{full} CFFTs.

This paper has three main contributions. First, we reduce both multiplicative and additive
complexities of \emph{partial CFFTs}, which compute only
\textbf{part} of the spectral components~\cite{Costa04,Fedorenko06},
based on their properties as well as our CSE algorithm in~\cite{Chen08a}. Our partial CFFTs have smaller complexities
than those in~\cite{Costa04}. Second, we propose \emph{dual
partial CFFTs}, where only a subset of temporal components are
nonzero, and reduce their complexities.
Finally, applying our partial and dual partial CFFTs, we reduce the
complexities of time- and transform-domain RS decoders
significantly.

\section{Partial and Dual Partial CFFTs}\label{sec:cfft}
We now consider CFFTs in two special cases. One special case is when
only a subset of frequency components are needed, and we refer to
such CFFTs as partial CFFTs following the convention in~\cite{Costa04,Fedorenko06}.
The other special case is when a subset of temporal components are
all zeros.
The two special cases can be viewed as dual to each other;
Thus, for the lack of a better term, we refer to CFFTs in the second
special case as dual partial CFFTs.

In a partial CFFT, some frequency components are not needed. Thus,
we first eliminate the rows corresponding to the unnecessary
frequency components from the post-addition matrices,
possibly resulting in all-zero columns. We then remove the
all-zero columns from the reduced post-addition matrices, as well as
the corresponding entries in $\boldsymbol{c}$ or
$\boldsymbol{c}^\ast$ and the corresponding rows from the
pre-addition matrices. For dual partial CFFTs, some temporal
components are zeros. Thus, we first remove the corresponding
columns in the pre-addition matrices, leading to
all-zero rows. We then remove the all-zero rows
from the reduced pre-addition matrices and the corresponding entries
in $\boldsymbol{c}$ or $\boldsymbol{c}^\ast$ as well as the
corresponding columns from the post-addition matrices.

It was shown~\cite{Chen08a} full SCFFTs and ICFFTs are equivalent in terms of complexities; Using a similar argument we can show that SCFFTs and ICFFTs are also equivalent in partial and dual partial DFTs.
In both special cases of CFFTs, removing rows or columns from pre- and
post-addition matrices leads to reduced additive complexities, and
eliminating entries in $\boldsymbol{c}$ or $\boldsymbol{c}^\ast$
results in reduced multiplicative complexities. Both multiplicative
and additive complexity reductions depend on the type of CFFTs. Note
that $\boldsymbol{P}^T$ and $\boldsymbol{P}$ are sparse, while
$\boldsymbol{AQ}$, $(\boldsymbol{A}'\boldsymbol{Q})^T$, and
$\boldsymbol{Q}^T\boldsymbol{A}^{-1}$ are not. Thus, removing a certain
number of rows or columns from $\boldsymbol{P}^T$
($\boldsymbol{P}$, respectively) leads to less significant
reductions in additive complexities than from $\boldsymbol{AQ}$
($(\boldsymbol{A}'\boldsymbol{Q})^T$ and
$\boldsymbol{Q}^T\boldsymbol{A}^{-1}$, respectively). On the other
hand, after removing some rows (columns, respectively), a reduced $\boldsymbol{P}^T$ ($\boldsymbol{P}$, respectively)
is more likely to have all-zero columns (rows, respectively) that eliminate entries in $\boldsymbol{c}$ or $\boldsymbol{c}^*$ than
$\boldsymbol{AQ}$ ($(\boldsymbol{A}'\boldsymbol{Q})^T$ and
$\boldsymbol{Q}^T\boldsymbol{A}^{-1}$, respectively).
Thus, partial
DCFFTs have higher multiplicative complexities but lower additive
complexities than partial SCFFTs/ICFFTs; similarly, dual partial
DCFFTs lead to lower multiplicative complexities but higher additive
complexities than dual partial SCFFTs/ICFFTs.

The savings in multiplicative complexities by partial SCFFTs/ICFFTs (dual partial DCFFTs, respectively)
are improved by considering different permutations of
$\boldsymbol{F}'$ and $\boldsymbol{F}''$ ($\boldsymbol{f}'$, respectively) while preserving all cyclotomic cosets. These permutations do not impact
$\boldsymbol{P}^T$ ($\boldsymbol{P}$, respectively) in a full CFFT, but by permuting $\boldsymbol{F}'$ and $\boldsymbol{F}''$ ($\boldsymbol{f}'$, respectively) the removed rows (columns, respectively) in $\boldsymbol{P}^T$ ($\boldsymbol{P}$, respectively)
result in more all-zero columns (rows, respectively) and thus achieve greater savings
in multiplicative complexities. This technique is equivalent to the rotation of normal bases in~\cite{Costa04}.

In addition to the complexity reduction techniques discussed above, which utilizes only the properties of the DFTs, we also apply our CSE algorithm~\cite{Chen08a} to further reduce the additive
complexities of both partial and dual partial CFFTs.

Partial CFFTs and their applications in syndrome computation were considered in~\cite{Costa04,Fedorenko06}, while dual partial CFFTs have not been considered in the literature to the best of our knowledge. In Section~\ref{sec:synd}, we compare the complexities of syndrome computation based on a variety of approaches, including our partial CFFTs and those in~\cite{Costa04}. We do not compare to~\cite{Fedorenko06} because no details were provided.

\section{Reduced-Complexity RS Decoders}\label{sec:rs}
Using full CFFTs \cite{Chen08a} as well as partial and dual partial CFFTs described above, we propose both time- and transform-domain RS decoders with reduced
complexities.

    \begin{table*}[htb]
        \centering
        \caption{Complexity of Syndrome Computation}
        \label{tab:pfft}
        \small\addtolength{\tabcolsep}{-3pt}
		\scalebox{
		\ifCLASSOPTIONdraftcls
        0.72
        \else
		0.85
		\fi}{
        \begin{tabular}{|c|c|c|c|c|c|c|c|c|c|c|c|c||c|c|c||c|c|c||c|}
                \hline
                \multirow{2}{*}{$(n, k)$} & \multicolumn{2}{c|}{Horner's rule} & \multicolumn{2}{c|}{\cite{Zakharova92}} & \multicolumn{2}{c|}{ICFFT \cite{Costa04}} & \multicolumn{2}{c|}{\cite{Jeng99}} & \multicolumn{2}{c|}{\cite{Lin07}} & \multicolumn{3}{c|}{Prime-factor \cite{Truong06a}} & \multicolumn{3}{c|}{Our SCFFT/ICFFT} & \multicolumn{3}{c|}{Our DCFFT}\\
                \cline{2-20}
                & Mult. & Add. & Mult. & Add. & Mult. & Add. & Mult. & Add. & Mult. & Add. & Mult. & Add. & Total & Mult. & Add. & Total & Mult. & Add. & Total\\
                \hline
                $(255, 223)$ & 7874 & 8128 & 167 & 5440 & 149 & 5046 & 8160 & 8128 & 3060 & 4998 & 852 & 1804 & 14584 & 149 & 3970 & 6205 & 586 & 2850 & 11640\\
                \hline
                $(511, 447)$ & 32130 & 32640 & - & - & - & - & 32704 & 32640 & 9888 & 17819 & 5265 & 7309 & 35496 & 345 & 16471 & 22336 & 1014 & 7904 & 25142\\
                \hline
                $(1023, 895)$ & 129794 & 130816 & - & - & - & - & 130944 & 130816 & 33620 & 73185 & 6785 & 15775 & 144690 & 824 & 60741 & 76397 & 2827 & 25118 & 78831\\
                \hline
        \end{tabular}
        }
    \end{table*}

\subsection{Syndrome Computation}\label{sec:synd}
We implement syndrome computation, which is
used in both time- and transform-domain decoders, with partial CFFTs.
For $(255, 223), (511, 447), (1023, 895)$ RS codes, which are
selected due to their
widespread applications \cite{Truong06a}, we compare the complexities of syndrome computation based on our
partial SCFFTs/ICFFTs with the complexities of syndrome computation
based on partial CFFTs in~\cite{Costa04} and other approaches such as
Horner's rule, Zakharova's algorithm~\cite{Zakharova92}, and
the prime-factor FFT~\cite{Truong06a} in Table~\ref{tab:pfft}.
The results for the length-255 RS code using Horner's rule as well as the
algorithms in~\cite{Zakharova92} and \cite{Costa04} are obtained
from~\cite{Costa04}; The results for RS codes of lengths 511 and
1023 using Horner's rule and the algorithm in~\cite{Lin07} are
reproduced from~\cite{Lin07}; the numbers of multiplications and
additions for the prime-factor FFT~\cite{Truong06a} are reproduced
from~\cite{Truong06a}.
In comparison to these approaches except the prime-factor FFT~\cite{Truong06a}, our partial SCFFTs/ICFFTs apparently require smaller complexities for syndrome computation.
To compare to the prime-factor FFT~\cite{Truong06a}, we use the metric for the
total complexities $N_{total} = (2m-1)N_{mult} + N_{add}$ as in~\cite{Chen08a}. Syndrome computation based on our
partial SCFFTs/ICFFTs requires smaller total complexities than those based on the prime-factor FFT~\cite{Truong06a}.
We provide the details of the syndrome computation for the $(255, 223)$ RS code based on our partial SCFFT in the appendix.

\subsection{Chien Search and Forney's Formula}\label{sec:chien}

For errors-only (errors-and-erasures, respectively) decoders,
the Chien search evaluates the error locator polynomial of degree at most $t$ (errata locator polynomial of degree at most $2t$, respectively) over all the elements of the underlying field; each root leads to one error (errata) location. This evaluation is essentially a DFT of a vector for which only first $t+1$ ($2t+1$, respectively) temporal components are not zeros. Note that the Chien search in errors-and-erasures decoders needs to evaluate only the error locator polynomial if it is available.
For errors-only (errors-and-erasures, respectively) decoders,
Forney's formula evaluates two polynomials: one is the
error (errata, respectively) evaluator polynomial $A(x)$ and the
other polynomial $x\tau'(x)$ is based on the derivative of error
(errata, respectively) locator polynomial $\tau(x)$. The degree of
the error (errata, respectively) evaluator polynomial is less than
$t$ ($2t$, respectively), while the degree of $x\tau'(x)$ is no more
than $t$  ($2t$, respectively).
Roughly half of the coefficients in $x\tau'(x)$
are zero. Given these information, the techniques
explained above for dual partial CFFTs are again applicable. For
simplicity, we assume errors-and-erasures decoders
henceforth, and our results can be easily extended to errors-only
decoders.

The errata locator polynomial $\tau(x)$ satisfies
$\tau(x)=\hat{\tau}_e(x^2)+x\hat{\tau}_o(x^2)$, where
$\hat{\tau}_e(x^2)$ and $x\hat{\tau}_o(x^2)$ consist of terms with
even and odd degrees, respectively. Note that $\hat{\tau}_e(x)$ and
$\hat{\tau}_o(x)$ have degrees at most $t$ and $t-1$, respectively.
It is easily verified that $\hat{\tau}_o(x^2)=\tau'(x)$ for
characteristic-$2$ fields.

While the Chien search evaluates $\tau(x)$ at all $n=2^m-1$ points,
Forney's formula evaluates $A(x)$ and $\tau'(x)$ at up to $2t$
errata locations. Thus, \textbf{given} the errata locations, the
evaluations of $A(x)$ and $\tau'(x)$ are DFTs, in which not only part
of temporal components are zeros but also only part of frequency
components are needed. Thus, the complexity reduction techniques for
both partial and dual partial CFFTs are applicable, and our CSE
algorithm can be applied. However, since the errata locations vary,
it is infeasible to minimize the computational complexities ``on
the fly.'' Thus, we also evaluate $A(x)$ and $\tau'(x)$ over all
$n=2^m-1$ points. Since $\tau'(x)$ is evaluated
over all the points, its evaluation is useful
for both the Chien search and Forney's formula. Thus, the Chien
search and Forney's formula are carried out jointly.

The evaluation of the $A(x)$ is directly implemented as a dual
partial CFFT. For any $\alpha \neq 0$ in GF$(2^m)$, we can either
obtain $\hat{\tau}_e(x^2)|_{x=\alpha}$ by dual partial CFFTs, or
first evaluate $\hat{\tau}_e(x)|_{x=\alpha}$ by dual partial CFFTs
and then obtain $\hat{\tau}_e(x^2)|_{x=\alpha}$ by properly
permuting the frequency components. Although $\hat{\tau}_e(x^2)$ and
$\hat{\tau}_e(x)$ have the same number of non-zero terms, the
non-zero terms of $\hat{\tau}_e(x)$ fall into fewer cosets than
those of $\tau_o(x)$, so its evaluation based on dual partial CFFTs
has smaller multiplicative and additive complexities.
Similar to our approach for $\hat{\tau}_e(x^2)$, we have two options to obtain $x\hat{\tau}_o(x^2)|_{x=\alpha}$. The
first option is treat $x\hat{\tau}_o(x^2)$ as a polynomial of degree
at most $2t-1$ and obtain $x\hat{\tau}_o(x^2)|_{x=\alpha}$ using
dual partial CFFTs.
The other option is to first compute $\hat{\tau}_o(x)|_{x=\alpha}$
using dual partial CFFTs, then obtain
$\hat{\tau}_o(x^2)|_{x=\alpha}$ by permutation, and finally compute
$x \hat{\tau}_o(x^2)|_{x=\alpha}$. For the latter option, similar to
the reason given above, the evaluation of $\hat{\tau}_o(x)$ based on
dual partial CFFTs requires fewer multiplications and additions than
that of $\hat{\tau}_o(x^2)$, although they have the same number of
non-zero terms. However, the latter option also requires $n$ extra
multiplications. Thus, the latter option has higher multiplicative
complexities but lower additive complexities as opposed to the former
option.

We present the computational complexities of combined Chien search
and Forney's formula for errors-and-erasures decoders based on our
dual partial CFFTs in Table~\ref{tab:forney}.
Note that to evaluate $\tau(x)$
at all $2t$ errata locations, $2t$ additions are needed; Also, $2t$
divisions are needed to compute the errata values in Forney's
formula; Both are accounted for in the rows marked by ``Misc.'' The
rows marked by ``Sum'' sum up the numbers of field operations required to
evaluate $A(x)$,
$\hat{\tau}_e(x^2)$, and $x\hat{\tau}_o(x^2)$, as well as $2t$ additions and
$2t$ divisions mentioned above.
As in Section~\ref{sec:synd}, the total complexities of each individual step and the sum are measured by the
metric in~\cite{Chen08a}, and we assume division has the same complexity as multiplication.
They are
presented in the columns marked by ``Total.''
The complexities of the two options for
evaluating $x\hat{\tau}_o(x^2)$ are both given; the $n$ extra
multiplications in the second option are shown in the second terms
of the summations. Due to the $n$ extra multiplications, for lengths
255 and 511 the first option has a smaller total complexity; for
length 1023, the second option has a smaller total complexity.
For evaluating $x\hat{\tau}_o(x^2)$, the
option with smaller \textbf{total} complexity is used.
\begin{table*}[htb]
    \centering
    \caption{Complexity of Combined Chien Search and Forney's Formula for errors-and-erasures decoders}
    \label{tab:forney}
    \small\addtolength{\tabcolsep}{-3pt}
    \scalebox{
	\ifCLASSOPTIONdraftcls
	0.67
	\else
	0.79
	\fi}{
    \begin{tabular}{|c|c|c|c|c|c||c|c|c|c||c|c|c|c||c|c|c|c||c|}
        \hline
        \multicolumn{3}{|c|}{\multirow{2}{*}{$(n, k)$}} & \multicolumn{4}{c|}{Our DCFFT} & \multicolumn{4}{c|}{Our SCFFT/ICFFT} & \multicolumn{4}{c|}{Horner's rule} & \multicolumn{4}{c|}{\cite{Jeng99}}\\
        \cline{4-19}
        \multicolumn{3}{|c|}{} & Mult. & Add. & Div. & Total & Mult. & Add. & Div. & Total & Mult. & Add. & Div. & Total & Mult. & Add. & Div. & Total\\
        \hline
        \multirow{6}{*}{(255, 223)} & \multicolumn{2}{c|}{$A(x)$} & 149 & 3226 & 0 & 5461 & 586 & 2628 & 0 & 11418 & 992 & 992 & 0 & 15872 & - & - & - & -\\
            \cline{2-19}
            & \multicolumn{2}{c|}{$\hat{\tau}_e(x^2)$} & 78 & 1828 & 0 & 2998 & 586 & 1990 & 0 & 10780 & 4080 & 4080 & 0 & 65280 & - & - & - & -\\
            \cline{2-19}
            & \multirow{2}{*}{$x\hat{\tau}_o(x^2)$} & Option 1 & 108 & 3096 & 0 & 4716 & 586 & 1970 & 0 & 10760 &  \multirow{2}{*}{4080} & \multirow{2}{*}{3825} & \multirow{2}{*}{0} & \multirow{2}{*}{65025} & \multirow{2}{*}{-} & \multirow{2}{*}{-} & \multirow{2}{*}{-} & \multirow{2}{*}{-}\\
            \cline{3-11}
            & & Option 2 & 75+255 & 1827 & 0 & 6777 & 586+255 & 1955 & 0 & 14570 & & & & & & & &\\
            \cline{2-19}
            & \multicolumn{2}{c|}{Misc} & 0 & 255 & 32 & 735 & 0 & 255 & 32 & 735 & 0 & 255 & 32 & 735 & - & - & - & -\\
            \cline{2-19}
            & \multicolumn{2}{c|}{Sum} & 335 & 8405 & 32 & 13910 & 1758 & 6843 & 32 & 33693 & 9152 & 9152 & 32 & 146912 & 15810 & 16575 & 32 & 254205\\
        \hline
        \multirow{6}{*}{(511, 447)} & \multicolumn{2}{c|}{$A(x)$} & 345 & 12791 & 0 & 18656 & 1014 & 7767 & 0 & 25005 & 4032 & 4032 & 0 & 72576 & - & - & - & -\\
            \cline{2-19}
            & \multicolumn{2}{c|}{$\hat{\tau}_e(x^2)$} & 177 & 7802 & 0 & 10811 & 1014 & 5299 & 0 & 22537 & 16352 & 16352 & 0 & 294336 & - & - & - & -\\
            \cline{2-19}
            & \multirow{2}{*}{$x\hat{\tau}_o(x^2)$} & Option 1 & 248 & 12533 & 0 & 16749 & 1014 & 5332 & 0 & 22570 &\multirow{2}{*}{16352} & \multirow{2}{*}{15841} & \multirow{2}{*}{0} & \multirow{2}{*}{293825} & \multirow{2}{*}{-} & \multirow{2}{*}{-} & \multirow{2}{*}{-} & \multirow{2}{*}{-}\\
            \cline{3-11}
            & & Option 2 & 176+511 & 7809 & 0 & 19488 & 1014+511 & 5243 & 0 & 31168 &  & & & & & & &\\
            \cline{2-19}
            & \multicolumn{2}{c|}{Misc} & 0 & 511 & 64 & 1599 & 0 & 511 & 64 & 1599 & 0 & 511 & 64 & 1599 & - & - & - & -\\
            \cline{2-19}
            & \multicolumn{2}{c|}{Sum} & 770 & 33637 & 64 & 47815 & 3042 & 18909 & 64 & 71711 & 36736 & 36736 & 64 & 662336 & 64386 & 65919 & 64 & 1161569\\
        \hline
        \multirow{6}{*}{(1023, 895)} & \multicolumn{2}{c|}{$A(x)$} & 824 & 52557 & 0 & 68213 & 2827 & 24806 & 0 & 78519 & 16256 & 16256 & 0 & 325120 & - & - & - & -\\
            \cline{2-19}
            \cline{3-11}
            & \multicolumn{2}{c|}{$\hat{\tau}_e(x^2)$} & 430 & 30294 & 0 & 38464 & 2827 & 16517 & 0 & 70230 & 65472 & 65472 & 0 & 1309440 & - & - & - & -\\
            \cline{2-19}
            & \multirow{2}{*}{$x\hat{\tau}_o(x^2)$} & Option 1 & 541 & 51655 & 0 & 61934 & 2827 & 16647 & 0 & 70360 & \multirow{2}{*}{65472} & \multirow{2}{*}{64449} & \multirow{2}{*}{0} & \multirow{2}{*}{1308417} & \multirow{2}{*}{-} & \multirow{2}{*}{-} & \multirow{2}{*}{-} & \multirow{2}{*}{-}\\
            \cline{3-11}
            & & Option 2 & 429+1023 & 30244 & 0 & 57832 & 2827+1023 & 16407 & 0 & 89557 & & & & & & & & \\
            \cline{2-19}
            & \multicolumn{2}{c|}{Misc} & 0 & 1023 & 128 & 3455 & 0 & 1023 & 128 & 3455 & 0 & 1023 & 128 & 3455 & - & - & - & -\\
            \cline{2-19}
            & \multicolumn{2}{c|}{Sum} & 2706 & 114118 & 128 & 167964 & 8481 & 58993 & 128 & 222564 & 147200 & 147200 & 128 & 2946432 & 259842 & 262911 & 128 & 5202341\\
        \hline
    \end{tabular}
    }
\end{table*}

The computational complexities
based on our dual partial DCFFTs are compared to the
complexities based on Horner's rule in Table~\ref{tab:forney}.
We also reproduce the complexities of the Chien search and Forney's formula in~\cite{Jeng99} from~\cite{Truong06a}.
The combined Chien search and Forney's formula
based on our partial dual CFFTs achieves significantly smaller
computational complexities than those based on Horner's rule and in~\cite{Jeng99}.
We do not compare with the approaches in~\cite{Fedorenko02,Lin07,Truong01} because their computational complexities are not available.

\subsection{Example}

We provide a simple example to illustrate syndrome computation and Chien search based on our CFFTs. For simplicity, let us consider errors-and-erasures decoding of a $(31, 25)$ cyclic RS code over $\GF(2^5)$ defined by the primitive polynomial $p(x) = x^5 + x^2 + 1$. The generator polynomial for the RS code is given by $g(x) = (x - 1) (x - \alpha) \cdots (x - \alpha^{5})$, where $\alpha$ is a root of $p(x)$. There are seven cyclotomic cosets over this field.

Suppose the received vector is $\boldsymbol{r} = (r_0, r_1, \dots,
r_{n - 1})$. To compute the syndromes $S_i = \sum_{j = 0}^{n - 1}
r_j \alpha^{ij}$ for $0 \le i \le 5$, it involves only four
cyclotomic cosets: $\{0\}, \{2, 4, 8, 16, 1\}$, $\{6, 12, 24, 17,
3\}, \{10, 20, 9, 18, 5\}$.
As explained in Section~\ref{sec:cfft}, we have rotated the cosets in this order to reduce multiplicative complexity.
We do not specify the other cosets since they are irrelevant to our purpose.
Using the normal basis $(\alpha^3, \alpha^6, \alpha^{12}, \alpha^{24}, \alpha^{17})$ and the length-5
convolution algorithm in~\cite{Blahut83}, we first construct a
full SCFFT $\boldsymbol{S}' =
\boldsymbol{P}^T (\boldsymbol{c} \cdot
(\boldsymbol{A}'\boldsymbol{Q})^T\boldsymbol{r}')$, in which $\boldsymbol{S}'$ and
$\boldsymbol{r}'$ are permuted versions of $\boldsymbol{S} = (S_0, S_1, \dotsc, S_{30})^T$ and
$\boldsymbol{r} = (r_0, r_1, \dotsc, r_{30})^T$, both ordered in the chosen cosets.
For these
four cosets, their $\boldsymbol{P}_i^T$ and $\boldsymbol{c}_i$ are
given by $\boldsymbol{P}_0^T = [1]$, $\boldsymbol{c}_0 = (1)$,
$\boldsymbol{c}_1 = \boldsymbol{c}_2 = \boldsymbol{c}_3=
(1, \alpha, \alpha^{25}, \alpha^7, \alpha^2, \alpha^{16},
\alpha^4, \alpha^{28}, \alpha^{14}, \alpha^{27})^T$,
and $$\boldsymbol{P}_1^T = \boldsymbol{P}_2^T = \boldsymbol{P}_3^T =
    \begin{bmatrix}
        1 & 1 & 0 & 1 & 1 & 0 & 1 & 0 & 0 & 1\\
        1 & 0 & 1 & 1 & 1 & 0 & 0 & 0 & 1 & 1\\
        1 & 1 & 1 & 1 & 1 & 1 & 1 & 1 & 1 & 0\\
        1 & 1 & 1 & 1 & 0 & 0 & 0 & 1 & 0 & 1\\
        1 & 1 & 1 & 0 & 1 & 1 & 0 & 0 & 0 & 1
    \end{bmatrix}.$$
Let $\boldsymbol{A}' = [\boldsymbol{A}'_0 \mid \boldsymbol{A}'_1
\mid \dots \mid \boldsymbol{A}'_6]$,
$\boldsymbol{A}'\boldsymbol{Q} = [\boldsymbol{A}'_0\boldsymbol{Q}_0
\mid \boldsymbol{A}'_1\boldsymbol{Q}_1 \mid \dots \mid
\boldsymbol{A}'_6 \boldsymbol{Q}_6]$.

In the coset $\{2, 4, 8, 16, 1\}$, we need only $\{2, 4, 1\}$, and thus we remove the third and fourth rows in $\boldsymbol{P}_1^T$, resulting in the eighth column being all-zero. So we strike out the column and save one more multiplication. Let $\boldsymbol{P}'^T_1$ denote the reduced matrix. Note that other orders of the coset cannot produce all-zero columns. This can also be achieved by rotating the normal basis.
Since we only need $\{3\}$ and $\{5\}$ in the cosets $\{6,12,24,17,3\}$ and $\{10,20,9,18,5\}$, respectively, we obtain $\boldsymbol{P}'^T_2 = \boldsymbol{P}'^T_3$ by keeping only the last row in $\boldsymbol{P}_2^T$ and striking out the all-zero fourth, seventh, eighth and ninth columns.
Correspondingly, we remove the eighth row from $\boldsymbol{Q}_1^T$ and $\boldsymbol{c}_1$. $\boldsymbol{Q}'^T_2 = \boldsymbol{Q}'^T_3$ and $\boldsymbol{c}_2' = \boldsymbol{c}_3'$ are given by removing the fourth, seventh, eighth, and ninth rows from $\boldsymbol{Q}^T_2$ and $\boldsymbol{c}_2$, respectively.
Hence the syndromes can be computed by a partial SCFFT as
$$
\begin{bmatrix} S_0\\ S_2\\ S_4\\ S_1\\ S_3\\ S_5 \end{bmatrix} =
    \begin{bmatrix}
        \boldsymbol{P}_0^T & & &\\
        & \boldsymbol{P}'^T_1 & &\\
        & & \boldsymbol{P}'^T_2 &\\
        & & & \boldsymbol{P}'^T_3&
    \end{bmatrix}
    \left(
    \begin{bmatrix} \boldsymbol{c}_0\\ \boldsymbol{c}_1'\\ \boldsymbol{c}_2'\\ \boldsymbol{c}_3'\end{bmatrix} \boldsymbol{\cdot}
        \begin{bmatrix}
            (\boldsymbol{A}'_0\boldsymbol{Q}_0)^T\\
            (\boldsymbol{A}'_1\boldsymbol{Q}'_1)^T\\
            (\boldsymbol{A}'_2\boldsymbol{Q}'_2)^T\\
            (\boldsymbol{A}'_3\boldsymbol{Q}'_3)^T\\
        \end{bmatrix}
        \boldsymbol{r}'\right).$$
       If there are all-zero columns in $\boldsymbol{Q}_1'^T, \boldsymbol{Q}'^T_2, \boldsymbol{Q}'^T_3$, we can strike out those columns and further remove corresponding rows from $\boldsymbol{A}'^T_i$'s.

	   In the Chien search, the errata locator polynomial $\tau(x) = \sum_{i=0}^6\tau_i x^i$ has degree up to six. So we need to use $\{6, 3\}$ for the third coset. Thus the Chien search can by done by a dual partial DCFFT
$$      \begin{bmatrix}
    \boldsymbol{A}'_0\boldsymbol{Q}_0 \mid \boldsymbol{A}'_1\boldsymbol{Q}'_1 \mid
    \boldsymbol{A}'_2\boldsymbol{Q}''_2 \mid \boldsymbol{A}'_3\boldsymbol{Q}'_3
        \end{bmatrix}
        \left(
        \begin{bmatrix} \boldsymbol{c}_0\\ \boldsymbol{c}_1'\\ \boldsymbol{c}_2''\\ \boldsymbol{c}_3'\end{bmatrix} \boldsymbol{\cdot}
\begin{bmatrix}
        \boldsymbol{P}_0 & & &\\
        & \boldsymbol{P}'_1 & &\\
        & & \boldsymbol{P}''_2 &\\
        & & & \boldsymbol{P}'_3\\
    \end{bmatrix}
    \begin{bmatrix} \tau_0\\ \tau_2\\ \tau_4\\ \tau_1\\ \tau_6\\ \tau_3\\ \tau_5\end{bmatrix}\right),
$$
where $\boldsymbol{P}''_2$ is obtained by keeping only the first and last columns of $\boldsymbol{P}_2$, $\boldsymbol{c}''_2$ and $\boldsymbol{Q}''_2$ are obtained by removing the eighth and ninth rows from $\boldsymbol{c}_2$ and the corresponding columns from $\boldsymbol{Q}_2$.

The Chien search can be split into evaluating $\tau_e(x)=\hat\tau_e(x^2)$ and $\tau_o(x)=x\hat\tau_o(x^2)$ to accommodate Forney's formula.
The direct evaluation of $\tau_o(x)$ can be carried out by
$$      \begin{bmatrix}
    \boldsymbol{A}'_1\boldsymbol{Q}''_1 \mid \boldsymbol{A}'_2\boldsymbol{Q}'_2 \mid
    \boldsymbol{A}'_3\boldsymbol{Q}'_3 \end{bmatrix}
        \left(
    \begin{bmatrix} \boldsymbol{c}_1''\\ \boldsymbol{c}_2'\\ \boldsymbol{c}_3'\end{bmatrix} \boldsymbol{\cdot}
\begin{bmatrix}
        \boldsymbol{P}''_1 & &\\
        & \boldsymbol{P}'_2 &\\
        & & \boldsymbol{P}'_3
    \end{bmatrix}
    \begin{bmatrix} \tau_1\\ \tau_3\\ \tau_5 \end{bmatrix}\right),
$$
where $\boldsymbol{Q}''_1 = \boldsymbol{Q}'_2, \boldsymbol{c}''_1 = \boldsymbol{c}'_2$, and $\boldsymbol{P}''_1 = \boldsymbol{P}'_2$. Alternatively, the evaluation of $\hat{\tau}_o(x^2)$ can be carried out by
$$      \begin{bmatrix}
    \boldsymbol{A}'_0\boldsymbol{Q}_0 \mid \boldsymbol{A}'_1\boldsymbol{Q}'''_1
        \end{bmatrix}
        \left(
    \begin{bmatrix} \boldsymbol{c}_0\\ \boldsymbol{c}_1'''\end{bmatrix} \boldsymbol{\cdot}
\begin{bmatrix}
        \boldsymbol{P}_0 &\\
        & \boldsymbol{P}'''_1
    \end{bmatrix}
    \begin{bmatrix} \hat{\tau}_{o,0}\\ \hat{\tau}_{o,2}\\ \hat{\tau}_{o,1}\end{bmatrix}\right),
$$
where $\boldsymbol{Q}'''_1 = \boldsymbol{Q}''_2, \boldsymbol{c}'''_1 = \boldsymbol{c}''_2 $, and $\boldsymbol{P}'''_1 = \boldsymbol{P}''_2$. Similarly, the evaluation of $\hat{\tau}_e(x^2)$ can be carried out by
$$      \begin{bmatrix}
    \boldsymbol{A}'_0\boldsymbol{Q}_0 \mid \boldsymbol{A}'_1\boldsymbol{Q}'''_1 \mid \boldsymbol{A}'_2\boldsymbol{Q}'_2
        \end{bmatrix}
        \left(
        \begin{bmatrix} \boldsymbol{c}_0\\ \boldsymbol{c}_1'''\\ \boldsymbol{c}_2'\end{bmatrix} \boldsymbol{\cdot}
\begin{bmatrix}
        \boldsymbol{P}_0 & &\\
        & \boldsymbol{P}'''_1 &\\
        & & \boldsymbol{P}'_2
    \end{bmatrix}
    \begin{bmatrix} \hat{\tau}_{e,0}\\ \hat{\tau}_{e,2}\\ \hat{\tau}_{e,1}\\ \hat{\tau}_{e,3} \end{bmatrix}\right).
$$

Then we can apply our CSE algorithm to these reduced pre- and post-addition matrices and furthur reduce the numbers of additions, but such details are omitted.
It is easy to see that the Chien search based on our partial dual CFFTs achieves significantly smaller computational complexities.
\subsection{Transform-Domain and Time-Domain RS Decoders}\label{sec:rrs}
Replacing the prime-factor FFT~\cite{Truong06a} by our CFFTs proposed above, we propose a transform-domain RS decoder with the following steps:
(T.1) Compute the syndromes by our partial SCFFT;
(T.2) Use the inverse-free BMA~\cite{Jeng99} to obtain the errata locator polynomial $\tau(x)$;
(T.3) Compute the remaining syndromes by recursive extension using $\tau(x)$;
(T.4) Compute the error vector by full CFFT of the syndrome vector. Finally, the
corrected codeword is obtained by adding the received vector and the
error vector.
Similarly, we propose a time-domain RS decoder with
the following steps: t.1 and t.2 are the same as T.1 and T.2;
(t.3) Compute the errata evaluator polynomial $A(x)$;
(t.4) Find the error locations and error values by applying our combined Chien search and Forney's formula based on dual partial DCFFTs to $\tau(x)$ and $A(x)$.

    \begin{table*}[htb]
        \centering
        \caption{Complexity of Transform-Domain and Time-Domain errors-and-erasures RS Decoders}
        \label{tab:ttrs}
        \small\addtolength{\tabcolsep}{-3pt}
		\scalebox{
		\ifCLASSOPTIONdraftcls
        0.85
		\else
        1
		\fi}{
        \begin{tabular}{|c|c|c|c|c||c|c|c|c||c|c|c|c||c|}
                \hline
        \multicolumn{2}{|c|}{\multirow{2}{*}{$(n, k)$}} & \multicolumn{4}{c|}{(255, 223)} & \multicolumn{4}{c|}{(511, 447)} & \multicolumn{4}{c|}{(1023, 895)}\\
                \cline{3-14}
                \multicolumn{2}{|c|}{} & Mult. & Add. & Div. & Total & Mult. & Add. & Div. & Total & Mult. & Add. & Div. & Total\\
                \hline
        \multirow{2}{*}{T.4 (Inverse Transform)} & \cite{Truong06a} & 1135 & 3887 & 0 & 20912 & 6516 & 17506 & 0 & 128278 & 5915 & 30547 & 0 & 142932\\
        \cline{2-14}
		& \cite{Chen08a} & 586 & 6736 & 0 & 15526 & 1014 & 23130 & 0 & 40368 & 2827 & 75360 & 0 & 129073\\
        \hline
        \multirow{2}{*}{T.1+T.4} & \cite{Truong06a} & 1987 & 5691 & 0 & 35496 & 11781 & 24815 & 0 & 225092 & 12700 & 46322 & 0 & 287622\\
        \cline{2-14}
        & Ours & 735 & 10706 & 0 & 21731 & 1359 & 39601 & 0 & 62704 & 3651 & 136101 & 0 & 205470\\
        \hline
        \multirow{2}{*}{t.1+t.4} & \cite{Jeng99} & 23970 & 24703 & 32 & 384733 & 97090 & 98559 & 64 & 1750177 & 390786 & 393727 & 128 & 7821093\\
        \cline{2-14}
            & Ours & 484 & 12375 & 32 & 20115 & 1115 & 50108 & 64 & 70151 & 3530 & 174859 & 128 & 244361\\
        \hline
        T.2, t.2 (BMA) & All & 353 & 288 & 0 & 5583 & 1217 & 1088 & 0 & 21777 & 4481 & 4224 & 0 & 89363\\
                \hline
        T.3 (Remaining syndromes) & Both & 7136 & 6913 & 0 & 113953 & 28608 & 28161 & 0 & 514497 & 114560 & 113665 & 0 & 2290305\\
        \hline
        \multirow{2}{*}{T.1+T.2+T.3+T.4} & \cite{Truong06a} & 9476 & 12892 & 0 & 155032 & 41606 & 54064 & 0 & 761366 & 131741 & 164211 & 0 & 2667290\\
        \cline{2-14}
        & Ours & 8224 & 17907 & 0 & 141267 & 31184 & 68850 & 0 & 598978 & 122692 & 253990 & 0 & 2585138\\
        \hline
        t.3 (Errata evaluator poly.) & Both & 1089 & 1024 & 0 & 17359 & 4225 & 4096 & 0 & 75921 & 16641 & 16384 & 0 & 332563\\
        \hline
        \multirow{2}{*}{t.1+t.2+t.3+t.4} & \cite{Jeng99} & 25412 & 26015 & 32 & 407675 & 102532 & 103743 & 64 & 1847875 & 411908 & 414335 & 128 & 8243019\\
        \cline{2-14}
        & Ours & 1926 & 12679 & 32 & 42049 & 6557 & 55292 & 64 & 167849 & 24652 & 195467 & 128 & 666287\\
        \hline
        \end{tabular}
        }
    \end{table*}
We compare the complexities of our time- and transform-domain RS decoders for $(255, 223)$, $(511, 447)$, and $(1023, 895)$ RS codes with those in~\cite{Jeng99} and~\cite{Truong06a} respectively in Table~\ref{tab:ttrs}.
We are aware of the vast
literature on RS decoding, and~\cite{Jeng99} and~\cite{Truong06a}
are compared here since their data are directly comparable. The
computational complexities of the time-domain decoder in~\cite{Jeng99} and the transform-domain decoder in~\cite{Truong06a}
are all reproduced from~\cite{Truong06a}. The complexities of T.4 are reproduced from~\cite[Table~I]{Chen08a}. Note that all the
computational complexities are for
errors-and-erasures decoders. The complexities of T.1/t.1 and t.4 are already presented in Tables~\ref{tab:pfft} and \ref{tab:forney}.

We first compare the overall complexities of our
transform-domain RS decoder with those in~\cite{Truong06a},
presented in the row marked by ``T.1+T.2+T.3+T.4.'' Clearly our
transform-domain decoder achieves smaller complexities. However,
this comparison is somewhat misleading since our decoder differs
from that in~\cite{Truong06a} only in T.1 and T.4. We further
compare the combined complexities of T.1 and T.4 of our
transform-domain decoder and that in~\cite{Truong06a}, presented in
the row marked by ``T.1+T.4.'' Here, the transform portion of our
decoder achieves complexity reductions of 39\%, 72\%, and 29\%.

For time-domain RS decoders, in comparison to the decoder considered
in~\cite{Jeng99}, the overall complexities of our RS decoder are
90\%, 91\%, and 92\% smaller. Again, since the focus of this paper is on
transformation, it is more meaningful to compare only the steps
using DFTs: t.1 and t.4. The sums of the total complexities of t.1
 and t.4 are presented separately in the row marked by
``t.1+t.4.'' It can be seen that the transformation portion of
our decoder achieves 95\%, 96\%, and 97\% complexity savings over that in~\cite{Jeng99} for the three RS codes, respectively.

Finally, based on our results,
time-domain decoders have smaller complexities than
transform-domain decoders. This conclusion is different from that in~\cite{Truong06a}. However, the conclusion in~\cite{Truong06a} is
based on the comparison of transform-domain decoder using FFT and
time-domain decoder without FFT. In our comparison, both decoders
use CFFTs.

In this paper, we assume that RS decoders are implemented by
integrated circuits, and each CFFT consists of combinational logic and requires no memory. Hence, we consider only the total complexity due to finite field operations above since they directly correspond to combinational logic. The total complexities in Tables~\ref{tab:pfft}, \ref{tab:forney}, and \ref{tab:ttrs} also assume that the maximum of received symbols are processed concurrently so as to increase throughput. Thus, reduced total complexities by CFFTs translate into smaller areas. However, decoders based on CFFTs have fixed and irregular adder trees
for pre- and post-additions,
and thus are less conducive to transformations that trade time (throughput) for area than decoders based on other approaches (for example, Horner's rule).

\section*{Acknowledgment}
The authors are grateful to Prof. P. Trifonov for providing details of CFFTs and Prof. P. D. Chen for valuable discussions.

\appendix[Partial SCFFT for Syndrome Computation in (255, 223) RS Codes]
We provide the details of the syndrome computation for the (255, 223) RS code based on our partial SCFFT.

First, we reorder the received vector $\boldsymbol{r}$ based on cosets:

$\boldsymbol{r}' = (r_{0}$, $r_{1}$, $r_{2}$, $r_{4}$, $r_{8}$, $r_{16}$, $r_{32}$, $r_{64}$, $r_{128}$, $r_{3}$, $r_{6}$, $r_{12}$, $r_{24}$, $r_{48}$, $r_{96}$, $r_{192}$, $r_{129}$, $r_{5}$, $r_{10}$, $r_{20}$, $r_{40}$, $r_{80}$, $r_{160}$, $r_{65}$, $r_{130}$, $r_{131}$, $r_{7}$, $r_{14}$, $r_{28}$, $r_{56}$, $r_{112}$, $r_{224}$, $r_{193}$, $r_{66}$, $r_{132}$, $r_{9}$, $r_{18}$, $r_{36}$, $r_{72}$, $r_{144}$, $r_{33}$, $r_{11}$, $r_{22}$, $r_{44}$, $r_{88}$, $r_{176}$, $r_{97}$, $r_{194}$, $r_{133}$, $r_{67}$, $r_{134}$, $r_{13}$, $r_{26}$, $r_{52}$, $r_{104}$, $r_{208}$, $r_{161}$, $r_{195}$, $r_{135}$, $r_{15}$, $r_{30}$, $r_{60}$, $r_{120}$, $r_{240}$, $r_{225}$, $r_{34}$, $r_{68}$, $r_{136}$, $r_{17}$, $r_{98}$, $r_{196}$, $r_{137}$, $r_{19}$, $r_{38}$, $r_{76}$, $r_{152}$, $r_{49}$, $r_{138}$, $r_{21}$, $r_{42}$, $r_{84}$, $r_{168}$, $r_{81}$, $r_{162}$, $r_{69}$, $r_{226}$, $r_{197}$, $r_{139}$, $r_{23}$, $r_{46}$, $r_{92}$, $r_{184}$, $r_{113}$, $r_{70}$, $r_{140}$, $r_{25}$, $r_{50}$, $r_{100}$, $r_{200}$, $r_{145}$, $r_{35}$, $r_{141}$, $r_{27}$, $r_{54}$, $r_{108}$, $r_{216}$, $r_{177}$, $r_{99}$, $r_{198}$, $r_{71}$, $r_{142}$, $r_{29}$, $r_{58}$, $r_{116}$, $r_{232}$, $r_{209}$, $r_{163}$, $r_{31}$, $r_{62}$, $r_{124}$, $r_{248}$, $r_{241}$, $r_{227}$, $r_{199}$, $r_{143}$, $r_{37}$, $r_{74}$, $r_{148}$, $r_{41}$, $r_{82}$, $r_{164}$, $r_{73}$, $r_{146}$, $r_{39}$, $r_{78}$, $r_{156}$, $r_{57}$, $r_{114}$, $r_{228}$, $r_{201}$, $r_{147}$, $r_{43}$, $r_{86}$, $r_{172}$, $r_{89}$, $r_{178}$, $r_{101}$, $r_{202}$, $r_{149}$, $r_{45}$, $r_{90}$, $r_{180}$, $r_{105}$, $r_{210}$, $r_{165}$, $r_{75}$, $r_{150}$, $r_{47}$, $r_{94}$, $r_{188}$, $r_{121}$, $r_{242}$, $r_{229}$, $r_{203}$, $r_{151}$, $r_{51}$, $r_{102}$, $r_{204}$, $r_{153}$, $r_{53}$, $r_{106}$, $r_{212}$, $r_{169}$, $r_{83}$, $r_{166}$, $r_{77}$, $r_{154}$, $r_{55}$, $r_{110}$, $r_{220}$, $r_{185}$, $r_{115}$, $r_{230}$, $r_{205}$, $r_{155}$, $r_{59}$, $r_{118}$, $r_{236}$, $r_{217}$, $r_{179}$, $r_{103}$, $r_{206}$, $r_{157}$, $r_{61}$, $r_{122}$, $r_{244}$, $r_{233}$, $r_{211}$, $r_{167}$, $r_{79}$, $r_{158}$, $r_{63}$, $r_{126}$, $r_{252}$, $r_{249}$, $r_{243}$, $r_{231}$, $r_{207}$, $r_{159}$, $r_{85}$, $r_{170}$, $r_{87}$, $r_{174}$, $r_{93}$, $r_{186}$, $r_{117}$, $r_{234}$, $r_{213}$, $r_{171}$, $r_{91}$, $r_{182}$, $r_{109}$, $r_{218}$, $r_{181}$, $r_{107}$, $r_{214}$, $r_{173}$, $r_{95}$, $r_{190}$, $r_{125}$, $r_{250}$, $r_{245}$, $r_{235}$, $r_{215}$, $r_{175}$, $r_{111}$, $r_{222}$, $r_{189}$, $r_{123}$, $r_{246}$, $r_{237}$, $r_{219}$, $r_{183}$, $r_{119}$, $r_{238}$, $r_{221}$, $r_{187}$, $r_{127}$, $r_{254}$, $r_{253}$, $r_{251}$, $r_{247}$, $r_{239}$, $r_{223}$, $r_{191}).$

Pre-additions (3793 additions): $\boldsymbol{p} = (\boldsymbol{A}\boldsymbol{Q})^T \boldsymbol{r'}.$

$	t_{2399} = r'_{150} + r'_{197},
	t_{2404} = r'_{228} + t_{2399},
	t_{2247} = r'_{189} + r'_{190},
	t_{2085} = r'_{92} + r'_{234},
	t_{2056} = r'_{78} + r'_{153},
	t_{1917} = r'_{33} + r'_{122},
	t_{1853} = r'_{20} + r'_{219},
	t_{1788} = r'_{34} + r'_{70},
	t_{1707} = r'_{83} + r'_{203},
	t_{2397} = r'_{152} + t_{1707},
	t_{1662} = r'_{35} + r'_{92},
	t_{1572} = r'_{8} + r'_{233},
	t_{1571} = r'_{1} + r'_{193},
	t_{1561} = r'_{19} + r'_{41},
	t_{1553} = r'_{227} + r'_{245},
	t_{1516} = r'_{82} + r'_{113},
	t_{1455} = r'_{54} + r'_{121},
	t_{1617} = r'_{153} + t_{1455},
	t_{1425} = r'_{50} + r'_{223},
	t_{1371} = r'_{222} + r'_{242},
	t_{1326} = r'_{29} + r'_{114},
	t_{1313} = r'_{152} + r'_{199},
	t_{1308} = r'_{122} + r'_{220},
	t_{1277} = r'_{25} + r'_{35},
	t_{1244} = r'_{232} + r'_{246},
	t_{1240} = r'_{168} + r'_{191},
	t_{1221} = r'_{18} + r'_{166},
	t_{1218} = r'_{105} + r'_{133},
	t_{1177} = r'_{21} + r'_{49},
	t_{1150} = r'_{7} + r'_{176},
	t_{1368} = r'_{151} + t_{1150},
	t_{1134} = r'_{9} + r'_{179},
	t_{1127} = r'_{23} + r'_{219},
	t_{1118} = r'_{3} + r'_{85},
	t_{1101} = r'_{60} + r'_{63},
	t_{1092} = r'_{172} + r'_{239},
	t_{1085} = r'_{183} + r'_{214},
	t_{1082} = r'_{37} + r'_{89},
	t_{1066} = r'_{2} + r'_{184},
	t_{1060} = r'_{83} + r'_{119},
	t_{1059} = r'_{210} + r'_{226},
	t_{1054} = r'_{169} + r'_{182},
	t_{1420} = r'_{122} + t_{1054},
	t_{1043} = r'_{141} + r'_{195},
	t_{1024} = r'_{97} + r'_{178},
	t_{1022} = r'_{22} + r'_{194},
	t_{1676} = r'_{249} + t_{1022},
	t_{1008} = r'_{17} + r'_{162},
	t_{1532} = r'_{145} + t_{1008},
	t_{1365} = r'_{160} + t_{1008},
	t_{1547} = r'_{243} + t_{1365},
	t_{999} = r'_{177} + r'_{181},
	t_{993} = r'_{188} + r'_{236},
	t_{983} = r'_{125} + r'_{131},
	t_{970} = r'_{47} + r'_{139},
	t_{957} = r'_{33} + r'_{34},
	t_{1310} = r'_{205} + t_{957},
	t_{1438} = r'_{129} + t_{1310},
	t_{950} = r'_{158} + r'_{217},
	t_{1041} = r'_{241} + t_{950},
	t_{948} = r'_{81} + r'_{248},
	t_{946} = r'_{48} + r'_{160},
	t_{940} = r'_{199} + r'_{254},
	t_{931} = r'_{197} + r'_{221},
	t_{1457} = r'_{178} + t_{931},
	t_{913} = r'_{39} + r'_{216},
	t_{2122} = r'_{199} + t_{913},
	t_{1263} = r'_{112} + t_{913},
	t_{908} = r'_{127} + r'_{149},
	t_{1020} = r'_{59} + t_{908},
	t_{907} = r'_{4} + r'_{171},
	t_{905} = r'_{55} + r'_{56},
	t_{1052} = r'_{118} + t_{905},
	t_{903} = r'_{150} + r'_{252},
	t_{987} = r'_{140} + t_{903},
	t_{901} = r'_{109} + r'_{121},
	t_{1179} = r'_{76} + t_{901},
	t_{892} = r'_{62} + r'_{84},
	t_{864} = r'_{32} + r'_{123},
	t_{862} = r'_{51} + r'_{180},
	t_{2150} = r'_{83} + t_{862},
	t_{1343} = t_{862} + t_{1024},
	t_{1039} = r'_{87} + t_{862},
	t_{858} = r'_{25} + r'_{89},
	t_{817} = r'_{157} + r'_{208},
	t_{1387} = r'_{14} + t_{817},
	t_{812} = r'_{129} + r'_{152},
	t_{811} = r'_{13} + r'_{222},
	t_{795} = r'_{91} + r'_{192},
	t_{1446} = r'_{224} + t_{795},
	t_{1276} = r'_{7} + t_{795},
	t_{934} = r'_{66} + t_{795},
	t_{793} = r'_{0} + r'_{155},
	t_{1193} = r'_{99} + t_{793},
	t_{788} = r'_{16} + r'_{101},
	t_{1230} = r'_{77} + t_{788},
	t_{785} = r'_{15} + r'_{217},
	t_{874} = r'_{77} + t_{785},
	t_{1189} = r'_{21} + t_{874},
	t_{783} = r'_{74} + r'_{115},
	t_{1204} = r'_{234} + t_{783},
	t_{779} = r'_{168} + r'_{210},
	t_{1056} = r'_{165} + t_{779},
	t_{774} = r'_{161} + r'_{196},
	t_{754} = r'_{46} + r'_{220},
	t_{1622} = r'_{164} + t_{754},
	t_{936} = r'_{162} + t_{754},
	t_{2082} = r'_{178} + t_{936},
	t_{748} = r'_{52} + r'_{141},
	t_{1564} = r'_{66} + t_{748},
	t_{977} = r'_{128} + t_{748},
	t_{742} = r'_{86} + r'_{178},
	t_{1133} = r'_{43} + t_{742},
	t_{1470} = r'_{174} + t_{1133},
	t_{737} = r'_{115} + r'_{119},
	t_{1996} = r'_{32} + t_{737},
	t_{1376} = r'_{135} + t_{737},
	t_{1149} = r'_{161} + t_{737},
	t_{735} = r'_{65} + r'_{243},
	t_{1488} = r'_{88} + t_{735},
	t_{846} = r'_{246} + t_{735},
	t_{732} = r'_{215} + r'_{223},
	t_{731} = r'_{49} + r'_{240},
	t_{965} = r'_{15} + t_{731},
	t_{1158} = r'_{198} + t_{965},
	t_{1669} = r'_{99} + t_{1158},
	t_{728} = r'_{185} + r'_{189},
	t_{720} = r'_{14} + r'_{108},
	t_{973} = r'_{204} + t_{720},
	t_{718} = r'_{92} + r'_{118},
	t_{1131} = r'_{145} + t_{718},
	t_{1690} = t_{1131} + t_{1572},
	t_{769} = r'_{130} + t_{718},
	t_{1620} = t_{769} + t_{858},
	t_{716} = r'_{146} + r'_{236},
	t_{2370} = r'_{160} + t_{716},
	t_{1050} = r'_{96} + t_{716},
	t_{2406} = r'_{55} + t_{1050},
	t_{713} = r'_{101} + r'_{212},
	t_{797} = r'_{150} + t_{713},
	t_{709} = r'_{3} + r'_{144},
	t_{841} = t_{709} + t_{774},
	t_{1272} = r'_{169} + t_{841},
	t_{707} = r'_{28} + r'_{114},
	t_{1412} = r'_{215} + t_{707},
	t_{1110} = r'_{225} + t_{707},
	t_{706} = r'_{165} + r'_{245},
	t_{1069} = t_{706} + t_{735},
	t_{2149} = r'_{61} + t_{1069},
	t_{705} = r'_{17} + r'_{234},
	t_{704} = r'_{25} + r'_{201},
	t_{825} = r'_{175} + t_{704},
	t_{1028} = t_{825} + t_{864},
	t_{1145} = t_{1028} + t_{1092},
	t_{701} = r'_{142} + r'_{198},
	t_{954} = r'_{44} + t_{701},
	t_{976} = r'_{172} + t_{954},
	t_{696} = r'_{203} + r'_{207},
	t_{1288} = r'_{251} + t_{696},
	t_{695} = r'_{24} + r'_{233},
	t_{1459} = r'_{252} + t_{695},
	t_{693} = r'_{69} + r'_{169},
	t_{889} = r'_{253} + t_{693},
	t_{688} = r'_{29} + r'_{85},
	t_{1036} = t_{688} + t_{705},
	t_{1114} = t_{709} + t_{1036},
	t_{2170} = r'_{48} + t_{1114},
	t_{1226} = t_{797} + t_{1114},
	t_{679} = r'_{120} + r'_{134},
	t_{678} = r'_{19} + r'_{81},
	t_{1713} = r'_{33} + t_{678},
	t_{2398} = t_{1713} + t_{2397},
	t_{1383} = t_{678} + t_{1376},
	t_{1528} = r'_{214} + t_{1383},
	t_{677} = r'_{228} + r'_{237},
	t_{1520} = r'_{213} + t_{677},
	t_{861} = r'_{100} + t_{677},
	t_{674} = r'_{90} + r'_{124},
	t_{842} = r'_{26} + t_{674},
	t_{664} = r'_{80} + r'_{172},
	t_{663} = r'_{157} + r'_{160},
	t_{1078} = r'_{144} + t_{663},
	t_{1672} = r'_{18} + t_{1078},
	t_{1299} = r'_{93} + t_{1078},
	t_{1402} = t_{1299} + t_{1326},
	t_{660} = r'_{10} + r'_{104},
	t_{1574} = t_{660} + t_{905},
	t_{1038} = t_{660} + t_{993},
	t_{1403} = r'_{12} + t_{1038},
	t_{759} = r'_{221} + t_{660},
	t_{659} = r'_{74} + r'_{239},
	t_{1344} = r'_{4} + t_{659},
	t_{1526} = r'_{236} + t_{1344},
	t_{658} = r'_{8} + r'_{175},
	t_{1021} = t_{658} + t_{679},
	t_{651} = r'_{84} + r'_{176},
	t_{647} = r'_{5} + r'_{214},
	t_{865} = t_{647} + t_{679},
	t_{1422} = r'_{146} + t_{865},
	t_{646} = r'_{9} + r'_{44},
	t_{1874} = t_{646} + t_{1272},
	t_{656} = r'_{159} + t_{646},
	t_{1530} = t_{656} + t_{1221},
	t_{644} = r'_{91} + r'_{117},
	t_{2363} = t_{644} + t_{1343},
	t_{1130} = r'_{107} + t_{644},
	t_{1926} = r'_{168} + t_{1130},
	t_{755} = r'_{102} + t_{644},
	t_{643} = r'_{38} + r'_{238},
	t_{828} = r'_{133} + t_{643},
	t_{1197} = r'_{14} + t_{828},
	t_{638} = r'_{11} + r'_{211},
	t_{2280} = t_{638} + t_{1056},
	t_{634} = r'_{158} + r'_{162},
	t_{632} = r'_{113} + r'_{235},
	t_{981} = r'_{251} + t_{632},
	t_{866} = r'_{68} + t_{632},
	t_{1664} = t_{866} + t_{1060},
	t_{627} = r'_{173} + r'_{249},
	t_{623} = r'_{30} + r'_{193},
	t_{804} = t_{623} + t_{674},
	t_{1517} = t_{804} + t_{1101},
	t_{2019} = t_{1517} + t_{1564},
	t_{1492} = r'_{147} + t_{804},
	t_{1503} = t_{977} + t_{1492},
	t_{1860} = r'_{11} + t_{1503},
	t_{622} = r'_{56} + r'_{73},
	t_{909} = r'_{170} + t_{622},
	t_{1760} = t_{731} + t_{909},
	t_{1049} = r'_{57} + t_{909},
	t_{1784} = r'_{6} + t_{1049},
	t_{770} = t_{622} + t_{705},
	t_{1037} = r'_{83} + t_{770},
	t_{1102} = r'_{141} + t_{1037},
	t_{1300} = t_{647} + t_{1102},
	t_{621} = r'_{40} + r'_{186},
	t_{1167} = t_{621} + t_{861},
	t_{914} = t_{621} + t_{664},
	t_{1529} = t_{647} + t_{914},
	t_{1769} = t_{1529} + t_{1574},
	t_{1227} = r'_{56} + t_{914},
	t_{619} = r'_{51} + r'_{148},
	t_{1400} = t_{619} + t_{720},
	t_{1007} = r'_{142} + t_{619},
	t_{618} = r'_{174} + r'_{250},
	t_{751} = r'_{137} + t_{618},
	t_{1033} = r'_{231} + t_{751},
	t_{615} = r'_{67} + r'_{167},
	t_{1274} = r'_{28} + t_{615},
	t_{614} = r'_{88} + r'_{126},
	t_{953} = r'_{31} + t_{614},
	t_{1494} = r'_{177} + t_{953},
	t_{745} = r'_{242} + t_{614},
	t_{1095} = r'_{77} + t_{745},
	t_{613} = r'_{54} + r'_{191},
	t_{951} = t_{613} + t_{658},
	t_{1443} = r'_{75} + t_{951},
	t_{768} = r'_{72} + t_{613},
	t_{1120} = t_{768} + t_{828},
	t_{607} = r'_{151} + r'_{155},
	t_{604} = r'_{39} + r'_{57},
	t_{2361} = r'_{242} + t_{604},
	t_{1144} = r'_{72} + t_{604},
	t_{1534} = t_{1144} + t_{1455},
	t_{2003} = t_{618} + t_{1534},
	t_{2010} = t_{1024} + t_{2003},
	t_{727} = r'_{79} + t_{604},
	t_{986} = t_{659} + t_{727},
	t_{603} = r'_{145} + r'_{213},
	t_{2083} = r'_{156} + t_{603},
	t_{654} = t_{603} + t_{627},
	t_{1139} = t_{654} + t_{948},
	t_{602} = r'_{105} + r'_{154},
	t_{2291} = t_{602} + t_{643},
	t_{1014} = r'_{208} + t_{602},
	t_{781} = r'_{209} + t_{602},
	t_{1486} = t_{732} + t_{781},
	t_{1569} = r'_{38} + t_{1486},
	t_{1654} = r'_{100} + t_{1569},
	t_{601} = r'_{70} + r'_{248},
	t_{897} = r'_{40} + t_{601},
	t_{1576} = r'_{55} + t_{897},
	t_{683} = r'_{96} + t_{601},
	t_{1186} = r'_{48} + t_{683},
	t_{600} = r'_{103} + r'_{107},
	t_{598} = r'_{170} + r'_{189},
	t_{853} = r'_{235} + t_{598},
	t_{1053} = r'_{42} + t_{853},
	t_{595} = r'_{122} + r'_{136},
	t_{1548} = t_{595} + t_{812},
	t_{762} = r'_{200} + t_{595},
	t_{1235} = r'_{73} + t_{762},
	t_{1137} = r'_{35} + t_{762},
	t_{1626} = t_{660} + t_{1137},
	t_{933} = r'_{137} + t_{762},
	t_{594} = r'_{132} + r'_{206},
	t_{1169} = r'_{244} + t_{594},
	t_{593} = r'_{109} + r'_{226},
	t_{1661} = t_{593} + t_{601},
	t_{1718} = r'_{247} + t_{1661},
	t_{612} = r'_{19} + t_{593},
	t_{1416} = t_{612} + t_{1134},
	t_{1790} = r'_{27} + t_{1416},
	t_{721} = r'_{171} + t_{612},
	t_{927} = r'_{143} + t_{721},
	t_{1575} = t_{927} + t_{1470},
	t_{1382} = r'_{107} + t_{927},
	t_{592} = r'_{22} + r'_{231},
	t_{670} = r'_{47} + t_{592},
	t_{1076} = t_{670} + t_{769},
	t_{1605} = t_{934} + t_{1076},
	t_{836} = t_{595} + t_{670},
	t_{1621} = r'_{198} + t_{836},
	t_{949} = r'_{237} + t_{836},
	t_{2052} = t_{949} + t_{1457},
	t_{591} = r'_{45} + r'_{219},
	t_{1969} = t_{591} + t_{1095},
	t_{815} = r'_{10} + t_{591},
	t_{1081} = r'_{153} + t_{815},
	t_{1493} = r'_{228} + t_{1081},
	t_{2027} = r'_{217} + t_{1493},
	t_{590} = r'_{49} + r'_{53},
	t_{588} = r'_{149} + r'_{195},
	t_{1121} = t_{588} + t_{817},
	t_{1225} = t_{618} + t_{1121},
	t_{1974} = t_{1225} + t_{1969},
	t_{673} = r'_{140} + t_{588},
	t_{587} = r'_{43} + r'_{225},
	t_{1862} = t_{587} + t_{1860},
	t_{1863} = t_{1066} + t_{1862},
	t_{1100} = r'_{78} + t_{587},
	t_{616} = r'_{59} + t_{587},
	t_{1338} = r'_{226} + t_{616},
	t_{810} = t_{616} + t_{688},
	t_{1433} = r'_{28} + t_{810},
	t_{583} = r'_{135} + r'_{139},
	t_{1129} = r'_{64} + t_{583},
	t_{582} = r'_{111} + r'_{123},
	t_{2156} = t_{582} + t_{751},
	t_{2158} = r'_{15} + t_{2156},
	t_{2161} = r'_{175} + t_{2158},
	t_{1108} = r'_{95} + t_{582},
	t_{1454} = t_{1014} + t_{1108},
	t_{712} = r'_{50} + t_{582},
	t_{2344} = r'_{34} + t_{712},
	t_{764} = r'_{76} + t_{712},
	t_{920} = r'_{6} + t_{764},
	t_{1533} = r'_{47} + t_{920},
	t_{581} = r'_{62} + r'_{183},
	t_{2276} = t_{581} + t_{1707},
	t_{682} = r'_{100} + t_{581},
	t_{888} = t_{656} + t_{682},
	t_{1501} = t_{788} + t_{888},
	t_{580} = r'_{94} + r'_{98},
	t_{1906} = t_{580} + t_{897},
	t_{576} = r'_{20} + r'_{229},
	t_{1685} = r'_{106} + t_{576},
	t_{633} = r'_{8} + t_{576},
	t_{1212} = t_{633} + t_{1053},
	t_{574} = r'_{1} + r'_{218},
	t_{1340} = t_{574} + t_{701},
	t_{1414} = t_{682} + t_{1340},
	t_{1198} = t_{574} + t_{1007},
	t_{1307} = t_{1198} + t_{1276},
	t_{1638} = t_{1149} + t_{1307},
	t_{572} = r'_{7} + r'_{197},
	t_{599} = r'_{188} + t_{572},
	t_{702} = r'_{216} + t_{599},
	t_{884} = r'_{34} + t_{702},
	t_{1023} = t_{663} + t_{884},
	t_{571} = r'_{93} + r'_{184},
	t_{890} = t_{571} + t_{768},
	t_{1164} = r'_{57} + t_{890},
	t_{681} = r'_{209} + t_{571},
	t_{947} = t_{681} + t_{940},
	t_{570} = r'_{58} + r'_{179},
	t_{1580} = t_{570} + t_{858},
	t_{859} = r'_{230} + t_{570},
	t_{1909} = t_{859} + t_{1561},
	t_{1248} = t_{859} + t_{1145},
	t_{569} = r'_{71} + r'_{75},
	t_{1877} = r'_{149} + t_{569},
	t_{1879} = t_{1874} + t_{1877},
	t_{1881} = t_{664} + t_{1879},
	t_{1105} = t_{569} + t_{634},
	t_{2339} = r'_{39} + t_{1105},
	t_{567} = r'_{36} + r'_{190},
	t_{2152} = t_{567} + t_{901},
	t_{730} = t_{567} + t_{651},
	t_{2086} = t_{730} + t_{755},
	t_{1124} = r'_{237} + t_{730},
	t_{1994} = r'_{59} + t_{1124},
	t_{564} = r'_{112} + r'_{138},
	t_{609} = r'_{164} + t_{564},
	t_{1637} = t_{609} + t_{949},
	t_{772} = r'_{131} + t_{609},
	t_{561} = r'_{12} + r'_{16},
	t_{845} = t_{561} + t_{600},
	t_{559} = r'_{78} + r'_{82},
	t_{558} = r'_{2} + r'_{143},
	t_{665} = r'_{194} + t_{558},
	t_{1087} = r'_{45} + t_{665},
	t_{1432} = t_{947} + t_{1087},
	t_{2215} = t_{1212} + t_{1432},
	t_{2218} = t_{764} + t_{2215},
	t_{1354} = r'_{60} + t_{1087},
	t_{1682} = r'_{65} + t_{1354},
	t_{2337} = t_{1020} + t_{1682},
	t_{771} = t_{665} + t_{678},
	t_{556} = r'_{95} + r'_{99},
	t_{553} = r'_{27} + r'_{31},
	t_{837} = t_{553} + t_{556},
	t_{552} = r'_{106} + r'_{128},
	t_{1040} = r'_{156} + t_{552},
	t_{1295} = r'_{244} + t_{1040},
	t_{630} = r'_{202} + t_{552},
	t_{1765} = r'_{21} + t_{630},
	t_{1772} = t_{908} + t_{1765},
	t_{1775} = r'_{47} + t_{1772},
	t_{767} = t_{619} + t_{630},
	t_{1972} = t_{558} + t_{767},
	t_{550} = r'_{60} + r'_{64},
	t_{2023} = t_{550} + t_{1414},
	t_{547} = r'_{32} + r'_{110},
	t_{1489} = t_{547} + t_{889},
	t_{1968} = t_{842} + t_{1489},
	t_{650} = r'_{205} + t_{547},
	t_{1849} = r'_{195} + t_{650},
	t_{860} = t_{650} + t_{673},
	t_{1170} = t_{860} + t_{1082},
	t_{519} = r'_{60} + r'_{82},
	t_{672} = r'_{37} + t_{519},
	t_{799} = r'_{63} + t_{672},
	t_{1332} = t_{664} + t_{799},
	t_{1560} = r'_{11} + t_{1332},
	t_{1607} = t_{1164} + t_{1560},
	t_{2062} = t_{1607} + t_{1622},
	t_{1080} = t_{759} + t_{799},
	t_{1581} = t_{594} + t_{1080},
	t_{1521} = t_{884} + t_{1080},
	t_{515} = r'_{34} + r'_{38},
	t_{514} = r'_{99} + r'_{251},
	t_{1257} = t_{514} + t_{934},
	t_{715} = r'_{163} + t_{514},
	t_{814} = t_{715} + t_{716},
	t_{510} = r'_{49} + r'_{75},
	t_{967} = t_{510} + t_{772},
	t_{1334} = t_{651} + t_{967},
	t_{1409} = t_{1167} + t_{1334},
	t_{1796} = t_{1409} + t_{1784},
	t_{509} = r'_{158} + r'_{189},
	t_{1667} = r'_{238} + t_{509},
	t_{1088} = r'_{116} + t_{509},
	t_{1419} = t_{591} + t_{1088},
	t_{508} = r'_{27} + r'_{107},
	t_{585} = r'_{224} + t_{508},
	t_{675} = r'_{129} + t_{585},
	t_{906} = t_{638} + t_{675},
	t_{740} = t_{598} + t_{675},
	t_{1562} = r'_{154} + t_{740},
	t_{944} = t_{740} + t_{815},
	t_{2173} = t_{944} + t_{1149},
	t_{506} = r'_{133} + r'_{137},
	t_{504} = r'_{236} + r'_{240},
	t_{502} = r'_{12} + r'_{155},
	t_{1044} = r'_{76} + t_{502},
	t_{1405} = r'_{87} + t_{1044},
	t_{499} = r'_{167} + r'_{168},
	t_{2091} = t_{499} + t_{613},
	t_{2093} = t_{1085} + t_{2091},
	t_{893} = r'_{26} + t_{499},
	t_{2270} = t_{893} + t_{1204},
	t_{1266} = t_{677} + t_{893},
	t_{498} = r'_{58} + r'_{62},
	t_{497} = r'_{149} + r'_{153},
	t_{494} = r'_{128} + r'_{132},
	t_{493} = r'_{136} + r'_{140},
	t_{492} = r'_{162} + r'_{185},
	t_{2318} = t_{492} + t_{1718},
	t_{2265} = t_{492} + t_{600},
	t_{807} = r'_{41} + t_{492},
	t_{1084} = t_{634} + t_{807},
	t_{549} = t_{492} + t_{509},
	t_{491} = r'_{59} + r'_{63},
	t_{982} = t_{491} + t_{846},
	t_{2125} = r'_{132} + t_{982},
	t_{490} = r'_{150} + r'_{154},
	t_{1384} = r'_{104} + t_{490},
	t_{1254} = t_{490} + t_{561},
	t_{577} = t_{490} + t_{506},
	t_{489} = r'_{204} + r'_{208},
	t_{488} = r'_{102} + r'_{106},
	t_{919} = r'_{205} + t_{488},
	t_{1122} = t_{919} + t_{1050},
	t_{1640} = t_{1122} + t_{1400},
	t_{2124} = t_{553} + t_{1640},
	t_{1062} = t_{643} + t_{919},
	t_{1703} = t_{1062} + t_{1526},
	t_{628} = t_{488} + t_{494},
	t_{1807} = t_{628} + t_{728},
	t_{487} = r'_{10} + r'_{14},
	t_{998} = t_{487} + t_{556},
	t_{773} = t_{487} + t_{489},
	t_{486} = r'_{180} + r'_{184},
	t_{1190} = r'_{24} + t_{486},
	t_{485} = r'_{95} + r'_{247},
	t_{1171} = r'_{52} + t_{485},
	t_{1885} = r'_{180} + t_{1171},
	t_{955} = t_{485} + t_{603},
	t_{1140} = t_{955} + t_{986},
	t_{1392} = r'_{223} + t_{1140},
	t_{1406} = t_{707} + t_{1392},
	t_{1551} = t_{502} + t_{1406},
	t_{791} = r'_{186} + t_{485},
	t_{1485} = t_{705} + t_{791},
	t_{1297} = r'_{50} + t_{791},
	t_{2135} = t_{489} + t_{1297},
	t_{2137} = t_{2125} + t_{2135},
	t_{2141} = t_{1085} + t_{2137},
	t_{2142} = t_{2124} + t_{2141},
	t_{596} = t_{485} + t_{514},
	t_{484} = r'_{141} + r'_{145},
	t_{482} = r'_{171} + r'_{175},
	t_{1243} = t_{482} + t_{576},
	t_{2095} = t_{1243} + t_{2093},
	t_{1971} = t_{1243} + t_{1637},
	t_{972} = t_{482} + t_{572},
	t_{481} = r'_{52} + r'_{56},
	t_{818} = t_{481} + t_{559},
	t_{2244} = t_{818} + t_{1534},
	t_{480} = r'_{227} + r'_{231},
	t_{1265} = t_{480} + t_{970},
	t_{938} = t_{480} + t_{683},
	t_{1270} = t_{892} + t_{938},
	t_{624} = t_{480} + t_{486},
	t_{1293} = t_{624} + t_{1105},
	t_{479} = r'_{109} + r'_{113},
	t_{2268} = t_{479} + t_{1257},
	t_{1011} = t_{479} + t_{588},
	t_{1496} = t_{1011} + t_{1382},
	t_{478} = r'_{36} + r'_{40},
	t_{1236} = r'_{200} + t_{478},
	t_{1395} = t_{1069} + t_{1236},
	t_{548} = t_{478} + t_{498},
	t_{1634} = t_{548} + t_{728},
	t_{1214} = t_{548} + t_{845},
	t_{1694} = t_{491} + t_{1214},
	t_{477} = r'_{213} + r'_{217},
	t_{575} = t_{477} + t_{484},
	t_{1658} = t_{575} + t_{1419},
	t_{476} = r'_{152} + r'_{156},
	t_{1468} = t_{476} + t_{506},
	t_{475} = r'_{67} + r'_{246},
	t_{2269} = r'_{224} + t_{475},
	t_{2274} = r'_{23} + t_{2269},
	t_{2275} = r'_{136} + t_{2274},
	t_{1602} = t_{475} + t_{781},
	t_{1336} = t_{475} + t_{1124},
	t_{803} = t_{475} + t_{754},
	t_{474} = r'_{193} + r'_{197},
	t_{473} = r'_{188} + r'_{192},
	t_{472} = r'_{16} + r'_{151},
	t_{962} = r'_{177} + t_{472},
	t_{1478} = t_{731} + t_{962},
	t_{645} = t_{472} + t_{594},
	t_{2335} = r'_{86} + t_{645},
	t_{2347} = t_{1520} + t_{2335},
	t_{2153} = t_{645} + t_{1414},
	t_{2162} = t_{2153} + t_{2161},
	t_{733} = r'_{227} + t_{645},
	t_{1715} = t_{733} + t_{1127},
	t_{922} = t_{650} + t_{733},
	t_{2073} = r'_{29} + t_{922},
	t_{1341} = t_{477} + t_{922},
	t_{565} = t_{472} + t_{502},
	t_{824} = t_{480} + t_{565},
	t_{471} = r'_{238} + r'_{242},
	t_{1347} = t_{471} + t_{508},
	t_{1566} = t_{903} + t_{1347},
	t_{1232} = t_{471} + t_{515},
	t_{2061} = t_{645} + t_{1232},
	t_{2063} = t_{2061} + t_{2062},
	t_{2065} = r'_{116} + t_{2063},
	t_{470} = r'_{219} + r'_{223},
	t_{469} = r'_{77} + r'_{81},
	t_{1161} = t_{469} + t_{715},
	t_{847} = t_{469} + t_{550},
	t_{694} = t_{469} + t_{583},
	t_{1219} = t_{638} + t_{694},
	t_{2189} = t_{1219} + t_{1626},
	t_{1222} = t_{672} + t_{1219},
	t_{468} = r'_{93} + r'_{97},
	t_{467} = r'_{17} + r'_{21},
	t_{2241} = t_{467} + t_{774},
	t_{1482} = r'_{5} + t_{467},
	t_{2261} = t_{634} + t_{1482},
	t_{466} = r'_{214} + r'_{218},
	t_{1508} = t_{466} + t_{580},
	t_{538} = t_{466} + t_{471},
	t_{714} = t_{504} + t_{538},
	t_{465} = r'_{110} + r'_{114},
	t_{464} = r'_{9} + r'_{13},
	t_{698} = r'_{181} + t_{464},
	t_{1187} = r'_{173} + t_{698},
	t_{1677} = t_{1187} + t_{1266},
	t_{463} = r'_{65} + r'_{244},
	t_{1851} = t_{463} + t_{1602},
	t_{1614} = r'_{42} + t_{463},
	t_{1911} = t_{1614} + t_{1906},
	t_{1913} = t_{616} + t_{1911},
	t_{462} = r'_{33} + r'_{37},
	t_{1539} = t_{462} + t_{1226},
	t_{1237} = t_{462} + t_{538},
	t_{461} = r'_{51} + r'_{55},
	t_{881} = r'_{194} + t_{461},
	t_{1380} = t_{881} + t_{1218},
	t_{1540} = t_{1300} + t_{1380},
	t_{1280} = r'_{68} + t_{881},
	t_{1910} = t_{1280} + t_{1402},
	t_{1912} = r'_{215} + t_{1910},
	t_{1920} = t_{1912} + t_{1913},
	t_{460} = r'_{139} + r'_{203},
	t_{1601} = r'_{13} + t_{460},
	t_{1372} = t_{460} + t_{906},
	t_{459} = r'_{98} + r'_{181},
	t_{964} = r'_{232} + t_{459},
	t_{1194} = r'_{249} + t_{964},
	t_{458} = r'_{237} + r'_{241},
	t_{1287} = t_{458} + t_{466},
	t_{925} = t_{458} + t_{481},
	t_{544} = t_{458} + t_{462},
	t_{2176} = t_{544} + t_{561},
	t_{1785} = t_{538} + t_{544},
	t_{952} = t_{471} + t_{544},
	t_{1635} = t_{952} + t_{1197},
	t_{457} = r'_{31} + r'_{103},
	t_{1479} = t_{457} + t_{1110},
	t_{830} = t_{457} + t_{755},
	t_{1570} = t_{751} + t_{830},
	t_{1720} = t_{815} + t_{1570},
	t_{912} = t_{812} + t_{830},
	t_{1498} = t_{460} + t_{912},
	t_{1373} = t_{912} + t_{993},
	t_{1702} = r'_{202} + t_{1373},
	t_{739} = t_{457} + t_{508},
	t_{1051} = t_{482} + t_{739},
	t_{589} = r'_{97} + t_{457},
	t_{456} = r'_{4} + r'_{8},
	t_{1148} = t_{456} + t_{706},
	t_{760} = t_{456} + t_{575},
	t_{1424} = r'_{39} + t_{760},
	t_{2271} = t_{1424} + t_{2265},
	t_{2273} = t_{1517} + t_{2271},
	t_{455} = r'_{64} + r'_{78},
	t_{1599} = r'_{110} + t_{455},
	t_{736} = t_{455} + t_{519},
	t_{597} = r'_{133} + t_{455},
	t_{1286} = t_{597} + t_{797},
	t_{988} = t_{597} + t_{811},
	t_{1999} = t_{614} + t_{988},
	t_{2000} = r'_{104} + t_{1999},
	t_{667} = r'_{254} + t_{597},
	t_{819} = r'_{71} + t_{667},
	t_{1172} = t_{819} + t_{1041},
	t_{1697} = t_{1172} + t_{1248},
	t_{935} = t_{623} + t_{819},
	t_{2130} = t_{732} + t_{935},
	t_{2131} = t_{772} + t_{2130},
	t_{1491} = r'_{36} + t_{935},
	t_{454} = r'_{70} + r'_{74},
	t_{843} = r'_{177} + t_{454},
	t_{1260} = r'_{27} + t_{843},
	t_{1116} = t_{698} + t_{843},
	t_{537} = t_{454} + t_{481},
	t_{766} = t_{537} + t_{596},
	t_{1656} = t_{766} + t_{1494},
	t_{1290} = t_{760} + t_{766},
	t_{453} = r'_{170} + r'_{174},
	t_{1659} = r'_{3} + t_{453},
	t_{586} = t_{453} + t_{473},
	t_{452} = r'_{80} + r'_{84},
	t_{610} = t_{452} + t_{487},
	t_{2031} = t_{610} + t_{952},
	t_{963} = t_{490} + t_{610},
	t_{451} = r'_{211} + r'_{215},
	t_{1027} = t_{451} + t_{714},
	t_{1947} = t_{963} + t_{1027},
	t_{566} = t_{451} + t_{491},
	t_{871} = t_{506} + t_{566},
	t_{1241} = t_{498} + t_{871},
	t_{746} = t_{544} + t_{566},
	t_{450} = r'_{50} + r'_{54},
	t_{1609} = t_{450} + t_{553},
	t_{857} = t_{450} + t_{783},
	t_{1604} = r'_{23} + t_{857},
	t_{1436} = r'_{53} + t_{857},
	t_{449} = r'_{26} + r'_{30},
	t_{1401} = t_{449} + t_{739},
	t_{711} = r'_{251} + t_{449},
	t_{1797} = t_{711} + t_{785},
	t_{448} = r'_{18} + r'_{22},
	t_{975} = t_{448} + t_{537},
	t_{1418} = r'_{200} + t_{975},
	t_{573} = t_{448} + t_{474},
	t_{838} = t_{573} + t_{624},
	t_{447} = r'_{144} + r'_{148},
	t_{1176} = t_{447} + t_{519},
	t_{446} = r'_{79} + r'_{83},
	t_{1946} = t_{446} + t_{550},
	t_{1958} = t_{1241} + t_{1946},
	t_{1321} = r'_{188} + t_{446},
	t_{2243} = t_{1321} + t_{1540},
	t_{445} = r'_{43} + r'_{47},
	t_{1348} = r'_{125} + t_{445},
	t_{1359} = t_{590} + t_{1348},
	t_{1636} = t_{1359} + t_{1528},
	t_{910} = t_{445} + t_{586},
	t_{1477} = r'_{51} + t_{910},
	t_{1259} = t_{480} + t_{910},
	t_{2123} = t_{1259} + t_{2122},
	t_{2128} = r'_{2} + t_{2123},
	t_{444} = r'_{57} + r'_{61},
	t_{959} = t_{444} + t_{847},
	t_{1421} = r'_{0} + t_{959},
	t_{443} = r'_{117} + r'_{121},
	t_{994} = t_{443} + t_{506},
	t_{1047} = r'_{60} + t_{994},
	t_{1070} = r'_{103} + t_{1047},
	t_{1378} = r'_{182} + t_{1070},
	t_{539} = t_{443} + t_{479},
	t_{442} = r'_{94} + r'_{177},
	t_{1546} = t_{442} + t_{497},
	t_{1495} = t_{442} + t_{1120},
	t_{1652} = t_{1129} + t_{1495},
	t_{848} = t_{442} + t_{793},
	t_{1519} = t_{848} + t_{1169},
	t_{1063} = t_{602} + t_{848},
	t_{1870} = r'_{94} + t_{1063},
	t_{655} = t_{442} + t_{589},
	t_{1030} = t_{655} + t_{771},
	t_{551} = t_{442} + t_{459},
	t_{1973} = t_{551} + t_{1546},
	t_{1675} = t_{551} + t_{695},
	t_{441} = r'_{87} + r'_{91},
	t_{2421} = t_{441} + t_{468},
	t_{776} = t_{441} + t_{633},
	t_{2340} = r'_{154} + t_{776},
	t_{2345} = r'_{94} + t_{2340},
	t_{1565} = t_{681} + t_{776},
	t_{1042} = r'_{171} + t_{776},
	t_{440} = r'_{72} + r'_{76},
	t_{1251} = t_{440} + t_{897},
	t_{2353} = r'_{254} + t_{1251},
	t_{1183} = t_{440} + t_{575},
	t_{560} = t_{440} + t_{468},
	t_{1178} = t_{560} + t_{998},
	t_{439} = r'_{186} + r'_{190},
	t_{438} = r'_{228} + r'_{232},
	t_{1721} = t_{438} + t_{1295},
	t_{1871} = t_{1297} + t_{1721},
	t_{437} = r'_{53} + r'_{71},
	t_{1168} = t_{437} + t_{612},
	t_{1147} = t_{437} + t_{767},
	t_{1732} = r'_{67} + t_{1147},
	t_{642} = t_{437} + t_{574},
	t_{2172} = t_{642} + t_{1037},
	t_{1094} = t_{630} + t_{642},
	t_{1317} = t_{842} + t_{1094},
	t_{1000} = t_{638} + t_{642},
	t_{541} = t_{437} + t_{510},
	t_{436} = r'_{104} + r'_{108},
	t_{1031} = t_{436} + t_{444},
	t_{1731} = t_{1031} + t_{1056},
	t_{780} = t_{436} + t_{497},
	t_{1717} = t_{549} + t_{780},
	t_{535} = t_{436} + t_{450},
	t_{1837} = t_{535} + t_{714},
	t_{939} = t_{497} + t_{535},
	t_{763} = t_{445} + t_{535},
	t_{1349} = t_{650} + t_{763},
	t_{1585} = t_{1230} + t_{1349},
	t_{435} = r'_{248} + r'_{252},
	t_{1071} = r'_{126} + t_{435},
	t_{2002} = t_{1071} + t_{1996},
	t_{2004} = t_{2000} + t_{2002},
	t_{2007} = t_{1994} + t_{2004},
	t_{652} = t_{435} + t_{452},
	t_{1573} = t_{652} + t_{983},
	t_{2320} = t_{538} + t_{1573},
	t_{434} = r'_{19} + r'_{23},
	t_{2105} = t_{434} + t_{1237},
	t_{527} = t_{434} + t_{438},
	t_{877} = t_{494} + t_{527},
	t_{1233} = t_{581} + t_{877},
	t_{700} = t_{527} + t_{539},
	t_{433} = r'_{202} + r'_{206},
	t_{1255} = t_{433} + t_{474},
	t_{2171} = t_{1095} + t_{1255},
	t_{2175} = t_{1601} + t_{2171},
	t_{2177} = t_{1043} + t_{2175},
	t_{657} = r'_{156} + t_{433},
	t_{1117} = t_{656} + t_{657},
	t_{832} = t_{510} + t_{657},
	t_{1115} = r'_{218} + t_{832},
	t_{584} = t_{433} + t_{493},
	t_{790} = t_{548} + t_{584},
	t_{1397} = t_{790} + t_{999},
	t_{432} = r'_{212} + r'_{216},
	t_{1017} = t_{432} + t_{535},
	t_{729} = t_{432} + t_{451},
	t_{2178} = r'_{250} + t_{729},
	t_{2179} = t_{2173} + t_{2178},
	t_{1502} = t_{552} + t_{729},
	t_{1098} = t_{466} + t_{729},
	t_{666} = t_{432} + t_{434},
	t_{1579} = t_{666} + t_{773},
	t_{606} = t_{432} + t_{515},
	t_{1026} = t_{606} + t_{736},
	t_{974} = t_{498} + t_{606},
	t_{431} = r'_{159} + r'_{163},
	t_{1234} = r'_{250} + t_{431},
	t_{2090} = t_{1234} + t_{2085},
	t_{1032} = r'_{61} + t_{431},
	t_{1350} = t_{1032} + t_{1038},
	t_{1445} = t_{1115} + t_{1350},
	t_{1559} = t_{1063} + t_{1445},
	t_{1716} = t_{982} + t_{1559},
	t_{896} = t_{431} + t_{489},
	t_{531} = t_{431} + t_{439},
	t_{1915} = r'_{235} + t_{531},
	t_{1919} = t_{1915} + t_{1917},
	t_{1922} = t_{455} + t_{1919},
	t_{747} = t_{531} + t_{541},
	t_{1535} = t_{739} + t_{747},
	t_{430} = r'_{111} + r'_{115},
	t_{995} = t_{430} + t_{700},
	t_{1610} = t_{476} + t_{995},
	t_{2364} = t_{1610} + t_{2152},
	t_{1984} = t_{1255} + t_{1610},
	t_{429} = r'_{165} + r'_{166},
	t_{2134} = t_{429} + t_{848},
	t_{937} = r'_{66} + t_{429},
	t_{1374} = t_{462} + t_{937},
	t_{511} = r'_{0} + t_{429},
	t_{1123} = r'_{138} + t_{511},
	t_{1588} = r'_{192} + t_{1123},
	t_{1758} = t_{858} + t_{1588},
	t_{1388} = t_{458} + t_{1123},
	t_{1886} = r'_{96} + t_{1388},
	t_{794} = r'_{116} + t_{511},
	t_{2220} = t_{794} + t_{962},
	t_{2228} = t_{1000} + t_{2220},
	t_{1089} = r'_{159} + t_{794},
	t_{1304} = t_{972} + t_{1089},
	t_{428} = r'_{160} + r'_{164},
	t_{1215} = t_{428} + t_{539},
	t_{1029} = t_{428} + t_{435},
	t_{427} = r'_{229} + r'_{233},
	t_{426} = r'_{3} + r'_{7},
	t_{1597} = t_{426} + t_{694},
	t_{2346} = t_{1597} + t_{2337},
	t_{2348} = t_{2345} + t_{2346},
	t_{2349} = t_{2347} + t_{2348},
	t_{2355} = t_{1571} + t_{2349},
	t_{787} = t_{426} + t_{476},
	t_{1009} = t_{494} + t_{787},
	t_{516} = t_{426} + t_{447},
	t_{2430} = t_{516} + t_{714},
	t_{1487} = t_{516} + t_{1017},
	t_{826} = t_{516} + t_{803},
	t_{425} = r'_{222} + r'_{226},
	t_{800} = t_{425} + t_{477},
	t_{1464} = t_{800} + t_{1387},
	t_{1426} = t_{800} + t_{1116},
	t_{424} = r'_{119} + r'_{123},
	t_{1698} = r'_{180} + t_{424},
	t_{1447} = t_{424} + t_{539},
	t_{1057} = t_{424} + t_{810},
	t_{806} = t_{424} + t_{440},
	t_{423} = r'_{1} + r'_{5},
	t_{2107} = t_{423} + t_{2105},
	t_{1281} = t_{423} + t_{1084},
	t_{1318} = r'_{31} + t_{1281},
	t_{1469} = r'_{164} + t_{1318},
	t_{738} = t_{423} + t_{590},
	t_{822} = t_{439} + t_{738},
	t_{422} = r'_{143} + r'_{147},
	t_{421} = r'_{221} + r'_{225},
	t_{2342} = t_{421} + t_{2339},
	t_{2350} = t_{1676} + t_{2342},
	t_{2351} = t_{2344} + t_{2350},
	t_{1386} = t_{421} + t_{874},
	t_{758} = t_{421} + t_{573},
	t_{1608} = t_{454} + t_{758},
	t_{1192} = r'_{9} + t_{758},
	t_{1497} = r'_{207} + t_{1192},
	t_{2216} = t_{1497} + t_{1547},
	t_{2224} = t_{1130} + t_{2216},
	t_{2225} = t_{727} + t_{2224},
	t_{529} = t_{421} + t_{445},
	t_{1954} = t_{489} + t_{529},
	t_{1512} = t_{529} + t_{700},
	t_{926} = t_{516} + t_{529},
	t_{1964} = t_{458} + t_{926},
	t_{420} = r'_{96} + r'_{100},
	t_{1003} = t_{420} + t_{654},
	t_{1199} = t_{1003} + t_{1060},
	t_{1206} = t_{1089} + t_{1199},
	t_{662} = t_{420} + t_{431},
	t_{902} = t_{504} + t_{662},
	t_{2192} = t_{425} + t_{902},
	t_{419} = r'_{86} + r'_{90},
	t_{2016} = t_{419} + t_{1519},
	t_{2018} = t_{592} + t_{2016},
	t_{2021} = t_{779} + t_{2018},
	t_{2022} = t_{1498} + t_{2021},
	t_{2024} = t_{2022} + t_{2023},
	t_{2028} = t_{2024} + t_{2027},
	t_{856} = r'_{139} + t_{419},
	t_{536} = t_{419} + t_{449},
	t_{1841} = t_{467} + t_{536},
	t_{418} = r'_{20} + r'_{24},
	t_{1333} = t_{418} + t_{936},
	t_{1356} = t_{674} + t_{1333},
	t_{1568} = r'_{231} + t_{1356},
	t_{1763} = t_{1551} + t_{1568},
	t_{1766} = t_{1546} + t_{1763},
	t_{699} = t_{418} + t_{443},
	t_{854} = t_{699} + t_{780},
	t_{2072} = t_{854} + t_{1421},
	t_{2078} = t_{806} + t_{2072},
	t_{671} = r'_{247} + t_{418},
	t_{1335} = t_{671} + t_{711},
	t_{534} = t_{418} + t_{427},
	t_{417} = r'_{134} + r'_{138},
	t_{1724} = r'_{26} + t_{417},
	t_{697} = t_{417} + t_{565},
	t_{1239} = t_{612} + t_{697},
	t_{1471} = t_{1089} + t_{1239},
	t_{530} = t_{417} + t_{476},
	t_{1945} = t_{530} + t_{696},
	t_{1951} = t_{577} + t_{1945},
	t_{611} = t_{464} + t_{530},
	t_{1513} = t_{577} + t_{611},
	t_{416} = r'_{172} + r'_{176},
	t_{1525} = t_{416} + t_{1485},
	t_{543} = t_{416} + t_{435},
	t_{882} = t_{452} + t_{543},
	t_{415} = r'_{194} + r'_{198},
	t_{1413} = t_{415} + t_{549},
	t_{1004} = t_{415} + t_{441},
	t_{2217} = t_{1004} + t_{1635},
	t_{1352} = r'_{98} + t_{1004},
	t_{414} = r'_{69} + r'_{73},
	t_{2264} = t_{414} + t_{1118},
	t_{2272} = t_{1352} + t_{2264},
	t_{2281} = t_{2272} + t_{2275},
	t_{2286} = t_{567} + t_{2281},
	t_{761} = t_{414} + t_{447},
	t_{1411} = t_{596} + t_{761},
	t_{918} = t_{580} + t_{761},
	t_{533} = t_{414} + t_{461},
	t_{2309} = t_{480} + t_{533},
	t_{1107} = r'_{110} + t_{533},
	t_{1342} = r'_{130} + t_{1107},
	t_{687} = t_{533} + t_{551},
	t_{1714} = t_{687} + t_{747},
	t_{929} = t_{465} + t_{687},
	t_{413} = r'_{157} + r'_{161},
	t_{1282} = t_{413} + t_{634},
	t_{1162} = t_{413} + t_{584},
	t_{775} = t_{413} + t_{470},
	t_{412} = r'_{66} + r'_{243},
	t_{1441} = t_{412} + t_{589},
	t_{1655} = r'_{186} + t_{1441},
	t_{1002} = r'_{65} + t_{412},
	t_{1249} = r'_{230} + t_{1002},
	t_{1630} = t_{1249} + t_{1446},
	t_{1789} = r'_{135} + t_{1630},
	t_{802} = t_{412} + t_{732},
	t_{1309} = r'_{33} + t_{802},
	t_{1005} = t_{567} + t_{802},
	t_{1370} = t_{572} + t_{1005},
	t_{503} = t_{412} + t_{463},
	t_{1699} = t_{503} + t_{826},
	t_{1223} = r'_{185} + t_{503},
	t_{528} = t_{499} + t_{503},
	t_{891} = t_{459} + t_{528},
	t_{1247} = t_{771} + t_{891},
	t_{1484} = t_{882} + t_{1247},
	t_{2379} = r'_{95} + t_{1484},
	t_{411} = r'_{35} + r'_{39},
	t_{1631} = t_{411} + t_{444},
	t_{629} = t_{411} + t_{497},
	t_{542} = t_{411} + t_{446},
	t_{1692} = t_{452} + t_{542},
	t_{1558} = t_{443} + t_{542},
	t_{1673} = r'_{12} + t_{1558},
	t_{1997} = t_{736} + t_{1673},
	t_{2001} = t_{1186} + t_{1997},
	t_{2009} = t_{2001} + t_{2007},
	t_{2011} = t_{2009} + t_{2010},
	t_{1557} = t_{542} + t_{1031},
	t_{2316} = t_{457} + t_{1557},
	t_{2324} = t_{425} + t_{2316},
	t_{2326} = t_{2241} + t_{2324},
	t_{2328} = t_{1131} + t_{2326},
	t_{835} = t_{542} + t_{549},
	t_{1977} = t_{835} + t_{1290},
	t_{410} = r'_{42} + r'_{46},
	t_{869} = r'_{147} + t_{410},
	t_{1680} = r'_{218} + t_{869},
	t_{1859} = t_{1443} + t_{1680},
	t_{1323} = t_{869} + t_{1161},
	t_{2242} = t_{1323} + t_{1519},
	t_{2252} = t_{907} + t_{2242},
	t_{1275} = t_{470} + t_{869},
	t_{409} = r'_{187} + r'_{191},
	t_{2266} = t_{409} + t_{1110},
	t_{2279} = t_{2266} + t_{2268},
	t_{2283} = t_{432} + t_{2279},
	t_{1181} = t_{409} + t_{533},
	t_{2168} = t_{761} + t_{1181},
	t_{668} = t_{409} + t_{416},
	t_{2336} = r'_{108} + t_{668},
	t_{1632} = t_{543} + t_{668},
	t_{792} = t_{542} + t_{668},
	t_{900} = t_{488} + t_{792},
	t_{1267} = t_{763} + t_{900},
	t_{540} = t_{409} + t_{428},
	t_{1213} = r'_{70} + t_{540},
	t_{1283} = t_{678} + t_{1213},
	t_{2020} = t_{1283} + t_{2019},
	t_{778} = t_{540} + t_{560},
	t_{408} = r'_{44} + r'_{48},
	t_{1143} = t_{408} + t_{671},
	t_{2352} = t_{460} + t_{1143},
	t_{2356} = t_{2351} + t_{2352},
	t_{1128} = t_{408} + t_{599},
	t_{2222} = r'_{169} + t_{1128},
	t_{1360} = t_{1021} + t_{1128},
	t_{1481} = r'_{73} + t_{1360},
	t_{1908} = r'_{237} + t_{1481},
	t_{684} = t_{408} + t_{438},
	t_{1152} = t_{684} + t_{715},
	t_{1061} = t_{611} + t_{684},
	t_{899} = t_{684} + t_{775},
	t_{1366} = t_{854} + t_{899},
	t_{532} = t_{408} + t_{425},
	t_{1523} = t_{532} + t_{582},
	t_{839} = t_{532} + t_{611},
	t_{407} = r'_{249} + r'_{253},
	t_{1678} = t_{407} + t_{468},
	t_{1914} = r'_{196} + t_{1678},
	t_{710} = t_{407} + t_{615},
	t_{2239} = r'_{227} + t_{710},
	t_{523} = t_{407} + t_{422},
	t_{978} = t_{461} + t_{523},
	t_{1933} = t_{530} + t_{978},
	t_{1428} = t_{978} + t_{1259},
	t_{945} = t_{450} + t_{523},
	t_{1653} = t_{945} + t_{1183},
	t_{637} = t_{469} + t_{523},
	t_{2079} = t_{490} + t_{637},
	t_{1200} = r'_{241} + t_{637},
	t_{984} = r'_{159} + t_{637},
	t_{1995} = t_{970} + t_{984},
	t_{2005} = r'_{221} + t_{1995},
	t_{1550} = t_{450} + t_{984},
	t_{2180} = t_{1386} + t_{1550},
	t_{406} = r'_{101} + r'_{105},
	t_{1068} = t_{406} + t_{695},
	t_{405} = r'_{127} + r'_{131},
	t_{1385} = r'_{69} + t_{405},
	t_{546} = t_{405} + t_{430},
	t_{404} = r'_{118} + r'_{122},
	t_{1627} = t_{404} + t_{1051},
	t_{1112} = t_{404} + t_{701},
	t_{1725} = t_{412} + t_{1112},
	t_{1556} = t_{679} + t_{1112},
	t_{725} = t_{404} + t_{465},
	t_{1567} = t_{725} + t_{939},
	t_{1111} = t_{405} + t_{725},
	t_{1591} = r'_{90} + t_{1111},
	t_{2136} = t_{1562} + t_{1591},
	t_{2139} = t_{2134} + t_{2136},
	t_{2144} = t_{2139} + t_{2142},
	t_{2145} = r'_{72} + t_{2144},
	t_{403} = r'_{41} + r'_{45},
	t_{1268} = t_{403} + t_{1040},
	t_{2219} = r'_{223} + t_{1268},
	t_{921} = t_{403} + t_{430},
	t_{1650} = t_{921} + t_{963},
	t_{2103} = t_{445} + t_{1650},
	t_{518} = t_{403} + t_{404},
	t_{887} = t_{470} + t_{518},
	t_{2385} = t_{486} + t_{887},
	t_{915} = t_{471} + t_{887},
	t_{741} = t_{427} + t_{518},
	t_{1767} = r'_{136} + t_{741},
	t_{1774} = t_{1766} + t_{1767},
	t_{402} = r'_{169} + r'_{173},
	t_{956} = t_{402} + t_{421},
	t_{813} = t_{402} + t_{417},
	t_{1612} = r'_{166} + t_{813},
	t_{886} = t_{619} + t_{813},
	t_{1125} = t_{616} + t_{886},
	t_{1301} = r'_{58} + t_{1125},
	t_{401} = r'_{135} + r'_{207},
	t_{1577} = t_{401} + t_{1068},
	t_{863} = t_{401} + t_{463},
	t_{1693} = t_{803} + t_{863},
	t_{1434} = r'_{245} + t_{863},
	t_{703} = r'_{187} + t_{401},
	t_{1582} = t_{621} + t_{703},
	t_{1660} = r'_{163} + t_{1582},
	t_{2392} = t_{1317} + t_{1660},
	t_{2395} = r'_{88} + t_{2392},
	t_{2410} = t_{2395} + t_{2406},
	t_{1119} = t_{703} + t_{1042},
	t_{1216} = r'_{108} + t_{1119},
	t_{1444} = t_{946} + t_{1216},
	t_{2278} = t_{1444} + t_{2273},
	t_{2284} = t_{2278} + t_{2280},
	t_{507} = t_{401} + t_{460},
	t_{966} = t_{456} + t_{507},
	t_{1104} = t_{409} + t_{966},
	t_{2195} = t_{1104} + t_{2192},
	t_{2201} = t_{1374} + t_{2195},
	t_{2054} = t_{1104} + t_{1139},
	t_{2055} = t_{1107} + t_{2054},
	t_{2060} = t_{1422} + t_{2055},
	t_{400} = r'_{230} + r'_{234},
	t_{620} = t_{400} + t_{467},
	t_{840} = t_{417} + t_{620},
	t_{517} = t_{400} + t_{424},
	t_{852} = t_{439} + t_{517},
	t_{1155} = t_{407} + t_{852},
	t_{649} = t_{467} + t_{517},
	t_{1642} = t_{649} + t_{760},
	t_{1987} = t_{1487} + t_{1642},
	t_{1229} = t_{404} + t_{649},
	t_{1511} = t_{546} + t_{1229},
	t_{399} = r'_{125} + r'_{129},
	t_{1339} = t_{399} + t_{806},
	t_{1930} = t_{852} + t_{1339},
	t_{1013} = t_{399} + t_{704},
	t_{1475} = t_{623} + t_{1013},
	t_{1315} = t_{1013} + t_{1095},
	t_{1362} = t_{933} + t_{1315},
	t_{717} = t_{399} + t_{478},
	t_{1555} = t_{696} + t_{717},
	t_{1935} = t_{780} + t_{1555},
	t_{1034} = t_{694} + t_{717},
	t_{2412} = t_{772} + t_{1034},
	t_{1515} = t_{532} + t_{1034},
	t_{680} = t_{399} + t_{415},
	t_{1381} = r'_{63} + t_{680},
	t_{1035} = t_{539} + t_{680},
	t_{2053} = t_{474} + t_{1035},
	t_{875} = t_{680} + t_{711},
	t_{557} = t_{399} + t_{465},
	t_{1246} = t_{557} + t_{1176},
	t_{1647} = t_{420} + t_{1246},
	t_{398} = r'_{28} + r'_{32},
	t_{1331} = r'_{218} + t_{398},
	t_{397} = r'_{68} + r'_{245},
	t_{923} = t_{397} + t_{429},
	t_{1404} = r'_{113} + t_{923},
	t_{1704} = r'_{143} + t_{1404},
	t_{1305} = t_{923} + t_{1052},
	t_{1648} = t_{1117} + t_{1305},
	t_{483} = r'_{210} + t_{397},
	t_{1430} = t_{483} + t_{484},
	t_{831} = r'_{182} + t_{483},
	t_{1584} = t_{569} + t_{831},
	t_{1067} = t_{831} + t_{864},
	t_{1273} = t_{693} + t_{1067},
	t_{2126} = t_{1273} + t_{1530},
	t_{2129} = t_{1516} + t_{2126},
	t_{2133} = t_{2129} + t_{2131},
	t_{2138} = t_{1119} + t_{2133},
	t_{2143} = t_{698} + t_{2138},
	t_{524} = r'_{209} + t_{483},
	t_{1872} = t_{524} + t_{720},
	t_{545} = t_{475} + t_{524},
	t_{1316} = t_{545} + t_{1194},
	t_{1611} = t_{1316} + t_{1418},
	t_{579} = t_{511} + t_{545},
	t_{608} = t_{528} + t_{579},
	t_{396} = r'_{195} + r'_{199},
	t_{1449} = t_{396} + t_{499},
	t_{941} = t_{396} + t_{414},
	t_{1554} = t_{557} + t_{941},
	t_{1086} = t_{482} + t_{941},
	t_{1681} = t_{704} + t_{1086},
	t_{2338} = t_{1681} + t_{2336},
	t_{2354} = t_{1342} + t_{2338},
	t_{2359} = t_{2354} + t_{2355},
	t_{2226} = t_{1653} + t_{1681},
	t_{2227} = t_{2218} + t_{2226},
	t_{1361} = t_{1086} + t_{1098},
	t_{1473} = t_{800} + t_{1361},
	t_{395} = r'_{250} + r'_{254},
	t_{2240} = t_{395} + t_{745},
	t_{2251} = t_{1632} + t_{2240},
	t_{2253} = t_{2251} + t_{2252},
	t_{2256} = t_{1664} + t_{2253},
	t_{1184} = t_{395} + t_{406},
	t_{1453} = t_{537} + t_{1184},
	t_{1156} = t_{395} + t_{416},
	t_{648} = t_{395} + t_{453},
	t_{1510} = t_{474} + t_{648},
	t_{1146} = t_{410} + t_{648},
	t_{1351} = r'_{188} + t_{1146},
	t_{2322} = t_{1351} + t_{2320},
	t_{2323} = r'_{51} + t_{2322},
	t_{928} = t_{540} + t_{648},
	t_{520} = t_{395} + t_{413},
	t_{1423} = r'_{75} + t_{520},
	t_{821} = t_{520} + t_{586},
	t_{1467} = t_{821} + t_{1026},
	t_{1458} = r'_{137} + t_{821},
	t_{2058} = t_{1458} + t_{2053},
	t_{2066} = t_{2058} + t_{2065},
	t_{2067} = t_{2052} + t_{2066},
	t_{1850} = t_{1270} + t_{1458},
	t_{1865} = t_{1850} + t_{1863},
	t_{394} = r'_{178} + r'_{182},
	t_{1979} = t_{394} + t_{1977},
	t_{782} = t_{394} + t_{532},
	t_{992} = t_{484} + t_{782},
	t_{1875} = r'_{27} + t_{992},
	t_{1882} = t_{1871} + t_{1875},
	t_{1883} = t_{1631} + t_{1882},
	t_{724} = t_{394} + t_{550},
	t_{1328} = t_{724} + t_{761},
	t_{961} = t_{672} + t_{724},
	t_{1587} = t_{961} + t_{987},
	t_{555} = t_{394} + t_{441},
	t_{979} = t_{555} + t_{837},
	t_{1624} = t_{450} + t_{979},
	t_{809} = t_{474} + t_{555},
	t_{1522} = t_{809} + t_{835},
	t_{2191} = t_{666} + t_{1522},
	t_{2199} = t_{548} + t_{2191},
	t_{1019} = t_{454} + t_{809},
	t_{393} = r'_{235} + r'_{239},
	t_{1729} = t_{393} + t_{456},
	t_{1103} = t_{393} + t_{856},
	t_{1298} = t_{558} + t_{1103},
	t_{878} = t_{393} + t_{629},
	t_{689} = t_{393} + t_{468},
	t_{796} = t_{529} + t_{689},
	t_{2197} = t_{524} + t_{796},
	t_{2202} = t_{2197} + t_{2201},
	t_{2203} = r'_{244} + t_{2202},
	t_{1595} = t_{488} + t_{796},
	t_{1898} = t_{845} + t_{1595},
	t_{1835} = t_{544} + t_{1595},
	t_{521} = t_{393} + t_{444},
	t_{911} = t_{428} + t_{521},
	t_{1719} = t_{911} + t_{1521},
	t_{1970} = r'_{211} + t_{1719},
	t_{1975} = t_{502} + t_{1970},
	t_{1976} = t_{1972} + t_{1975},
	t_{1978} = t_{1971} + t_{1976},
	t_{1980} = t_{1974} + t_{1978},
	t_{1981} = t_{1968} + t_{1980},
	t_{1982} = t_{1979} + t_{1981},
	t_{1983} = t_{1973} + t_{1982},
	t_{60} = t_{890} + t_{1983},
	t_{808} = t_{60} + t_{570},
	t_{930} = t_{462} + t_{911},
	t_{1065} = t_{652} + t_{930},
	t_{669} = t_{507} + t_{521},
	t_{2325} = t_{669} + t_{1699},
	t_{2331} = t_{1130} + t_{2325},
	t_{1641} = t_{669} + t_{1039},
	t_{1079} = t_{479} + t_{669},
	t_{883} = t_{491} + t_{669},
	t_{392} = r'_{179} + r'_{183},
	t_{1330} = t_{392} + t_{552},
	t_{851} = t_{392} + t_{504},
	t_{1895} = t_{851} + t_{1513},
	t_{1596} = t_{560} + t_{851},
	t_{1957} = t_{1426} + t_{1596},
	t_{1083} = t_{668} + t_{851},
	t_{1880} = t_{1083} + t_{1331},
	t_{1888} = t_{1605} + t_{1880},
	t_{1889} = t_{1883} + t_{1888},
	t_{526} = t_{392} + t_{420},
	t_{868} = t_{449} + t_{526},
	t_{1452} = t_{443} + t_{868},
	t_{719} = t_{526} + t_{543},
	t_{1753} = t_{719} + t_{1065},
	t_{1537} = t_{548} + t_{719},
	t_{1182} = t_{418} + t_{719},
	t_{391} = r'_{201} + r'_{205},
	t_{1174} = t_{391} + t_{847},
	t_{872} = t_{391} + t_{464},
	t_{2394} = t_{872} + t_{901},
	t_{2405} = t_{2394} + t_{2398},
	t_{1531} = t_{714} + t_{872},
	t_{1153} = r'_{216} + t_{872},
	t_{1099} = t_{493} + t_{872},
	t_{743} = t_{391} + t_{448},
	t_{1450} = t_{609} + t_{743},
	t_{2254} = t_{1450} + t_{2247},
	t_{2258} = t_{1153} + t_{2254},
	t_{1357} = t_{666} + t_{743},
	t_{1686} = t_{775} + t_{1357},
	t_{1097} = t_{743} + t_{814},
	t_{1852} = t_{1097} + t_{1849},
	t_{1857} = t_{918} + t_{1852},
	t_{1514} = t_{1097} + t_{1298},
	t_{522} = t_{391} + t_{406},
	t_{1695} = t_{522} + t_{1222},
	t_{2157} = t_{1695} + t_{2150},
	t_{2159} = t_{2149} + t_{2157},
	t_{2160} = t_{655} + t_{2159},
	t_{2166} = t_{710} + t_{2160},
	t_{867} = t_{522} + t_{559},
	t_{980} = t_{583} + t_{867},
	t_{2116} = t_{436} + t_{980},
	t_{2117} = t_{606} + t_{2116},
	t_{1988} = t_{839} + t_{980},
	t_{636} = t_{522} + t_{577},
	t_{1986} = t_{494} + t_{636},
	t_{1989} = t_{882} + t_{1986},
	t_{1990} = t_{928} + t_{1989},
	t_{1663} = t_{425} + t_{636},
	t_{855} = t_{493} + t_{636},
	t_{1992} = t_{855} + t_{1017},
	t_{1618} = t_{773} + t_{855},
	t_{390} = r'_{220} + r'_{224},
	t_{1073} = r'_{196} + t_{390},
	t_{1598} = t_{1073} + t_{1244},
	t_{2233} = t_{1193} + t_{1598},
	t_{2234} = t_{2228} + t_{2233},
	t_{2089} = t_{1598} + t_{2083},
	t_{639} = t_{390} + t_{410},
	t_{2368} = t_{639} + t_{1168},
	t_{2371} = t_{2363} + t_{2368},
	t_{1839} = t_{639} + t_{1835},
	t_{1843} = t_{1837} + t_{1839},
	t_{1399} = t_{639} + t_{747},
	t_{1683} = t_{945} + t_{1399},
	t_{960} = t_{541} + t_{639},
	t_{2306} = t_{596} + t_{960},
	t_{1625} = t_{877} + t_{960},
	t_{389} = r'_{126} + r'_{130},
	t_{1759} = t_{389} + t_{859},
	t_{1761} = r'_{208} + t_{1759},
	t_{1762} = t_{1760} + t_{1761},
	t_{1768} = t_{1758} + t_{1762},
	t_{1771} = t_{1768} + t_{1769},
	t_{1773} = t_{503} + t_{1771},
	t_{1776} = t_{1773} + t_{1775},
	t_{1778} = t_{582} + t_{1776},
	t_{1010} = t_{389} + t_{656},
	t_{2071} = t_{792} + t_{1010},
	t_{2074} = t_{709} + t_{2071},
	t_{2081} = t_{2074} + t_{2079},
	t_{2084} = t_{2081} + t_{2082},
	t_{2088} = r'_{94} + t_{2084},
	t_{990} = t_{389} + t_{517},
	t_{2388} = t_{990} + t_{2385},
	t_{500} = t_{389} + t_{396},
	t_{723} = t_{489} + t_{500},
	t_{2296} = t_{723} + t_{867},
	t_{1398} = t_{549} + t_{723},
	t_{968} = r'_{238} + t_{723},
	t_{1242} = t_{659} + t_{968},
	t_{2403} = t_{627} + t_{1242},
	t_{640} = t_{500} + t_{534},
	t_{1733} = t_{640} + t_{1608},
	t_{1465} = t_{620} + t_{640},
	t_{388} = r'_{142} + r'_{146},
	t_{2401} = r'_{39} + t_{388},
	t_{1780} = t_{388} + t_{1508},
	t_{997} = t_{388} + t_{396},
	t_{1320} = t_{861} + t_{997},
	t_{1583} = r'_{126} + t_{1320},
	t_{756} = t_{388} + t_{569},
	t_{1480} = t_{756} + t_{1341},
	t_{1303} = t_{540} + t_{756},
	t_{1649} = t_{566} + t_{1303},
	t_{1090} = t_{662} + t_{756},
	t_{1619} = t_{918} + t_{1090},
	t_{722} = t_{388} + t_{553},
	t_{2075} = r'_{166} + t_{722},
	t_{2080} = t_{1654} + t_{2075},
	t_{2087} = t_{2078} + t_{2080},
	t_{2094} = t_{2087} + t_{2088},
	t_{2097} = t_{2094} + t_{2095},
	t_{1329} = t_{425} + t_{722},
	t_{1563} = t_{564} + t_{1329},
	t_{924} = t_{584} + t_{722},
	t_{1670} = t_{561} + t_{924},
	t_{505} = t_{388} + t_{423},
	t_{2427} = t_{505} + t_{1467},
	t_{2428} = t_{1714} + t_{2427},
	t_{1045} = t_{505} + t_{628},
	t_{631} = t_{422} + t_{505},
	t_{387} = r'_{25} + r'_{29},
	t_{1838} = t_{387} + t_{481},
	t_{1840} = t_{479} + t_{1838},
	t_{1141} = t_{387} + t_{610},
	t_{1460} = t_{473} + t_{1141},
	t_{820} = t_{387} + t_{703},
	t_{1324} = r'_{229} + t_{820},
	t_{1456} = t_{1324} + t_{1362},
	t_{1231} = r'_{102} + t_{820},
	t_{1346} = t_{471} + t_{1231},
	t_{708} = t_{387} + t_{520},
	t_{1730} = t_{708} + t_{1663},
	t_{827} = t_{708} + t_{728},
	t_{1644} = t_{724} + t_{827},
	t_{2104} = t_{425} + t_{1644},
	t_{2106} = t_{2103} + t_{2104},
	t_{2112} = t_{416} + t_{2106},
	t_{386} = r'_{112} + r'_{116},
	t_{870} = t_{386} + t_{546},
	t_{1592} = t_{736} + t_{870},
	t_{1252} = r'_{22} + t_{870},
	t_{1220} = t_{402} + t_{870},
	t_{1687} = t_{1220} + t_{1267},
	t_{753} = t_{386} + t_{398},
	t_{2194} = t_{738} + t_{753},
	t_{1018} = t_{666} + t_{753},
	t_{385} = r'_{196} + r'_{200},
	t_{1736} = t_{385} + t_{553},
	t_{1427} = t_{385} + t_{778},
	t_{1209} = t_{385} + t_{824},
	t_{895} = t_{385} + t_{639},
	t_{1302} = t_{700} + t_{895},
	t_{2420} = t_{687} + t_{1302},
	t_{1210} = t_{814} + t_{895},
	t_{786} = t_{385} + t_{655},
	t_{1278} = t_{742} + t_{786},
	t_{1394} = t_{427} + t_{1278},
	t_{525} = t_{385} + t_{415},
	t_{1524} = t_{525} + t_{606},
	t_{2210} = t_{413} + t_{1524},
	t_{1196} = t_{525} + t_{929},
	t_{932} = t_{525} + t_{689},
	t_{1689} = t_{932} + t_{990},
	t_{2034} = t_{531} + t_{1689},
	t_{2035} = t_{902} + t_{2034},
	t_{686} = t_{406} + t_{525},
	t_{1782} = t_{686} + t_{1475},
	t_{1783} = t_{477} + t_{1782},
	t_{1258} = t_{586} + t_{686},
	t_{1269} = t_{422} + t_{1258},
	t_{2127} = t_{454} + t_{1269},
	t_{2132} = t_{1071} + t_{2127},
	t_{2140} = t_{2128} + t_{2132},
	t_{2146} = t_{1189} + t_{2140},
	t_{2147} = t_{2145} + t_{2146},
	t_{2148} = t_{2143} + t_{2147},
	t_{384} = r'_{120} + r'_{124},
	t_{1750} = t_{384} + t_{1683},
	t_{1752} = t_{1045} + t_{1750},
	t_{844} = t_{384} + t_{634},
	t_{1684} = t_{844} + t_{1608},
	t_{752} = t_{384} + t_{580},
	t_{1735} = t_{752} + t_{1733},
	t_{1737} = t_{728} + t_{1735},
	t_{1743} = t_{575} + t_{1737},
	t_{1188} = t_{453} + t_{752},
	t_{512} = t_{384} + t_{386},
	t_{1600} = t_{512} + t_{787},
	t_{1195} = t_{512} + t_{681},
	t_{798} = t_{441} + t_{512},
	t_{1390} = t_{662} + t_{798},
	t_{1379} = t_{391} + t_{798},
	t_{2208} = t_{1379} + t_{1555},
	t_{2151} = t_{838} + t_{1379},
	t_{2155} = r'_{155} + t_{2151},
	t_{2163} = t_{2155} + t_{2162},
	t_{2165} = t_{1125} + t_{2163},
	t_{2167} = t_{2165} + t_{2166},
	t_{784} = t_{512} + t_{556},
	t_{1389} = t_{784} + t_{1170},
	t_{2248} = t_{1389} + t_{2239},
	t_{1157} = t_{546} + t_{784},
	t_{1505} = t_{915} + t_{1157},
	t_{625} = t_{512} + t_{536},
	t_{2169} = t_{625} + t_{818},
	t_{2174} = t_{2169} + t_{2172},
	t_{2181} = t_{2170} + t_{2174},
	t_{2182} = t_{2177} + t_{2181},
	t_{2183} = t_{2180} + t_{2182},
	t_{2184} = t_{2168} + t_{2183},
	t_{1628} = t_{525} + t_{625},
	t_{1312} = t_{625} + t_{628},
	t_{1606} = t_{680} + t_{1312},
	t_{61} = t_{573} + t_{669} + t_{824} + t_{839} + t_{1051} + t_{1255} + t_{1290} + t_{1512} + t_{1522} + t_{1606},
	t_{1415} = t_{61} + t_{1098},
	t_{2193} = t_{407} + t_{1415},
	t_{1055} = t_{515} + t_{625},
	t_{2187} = t_{998} + t_{1055},
	t_{1337} = t_{397} + t_{1055},
	t_{2321} = t_{1337} + t_{2318},
	t_{2327} = t_{2261} + t_{2321},
	t_{2329} = t_{2323} + t_{2327},
	t_{2332} = t_{2329} + t_{2331},
	t_{383} = r'_{88} + r'_{92},
	t_{2389} = t_{383} + t_{409},
	t_{1764} = t_{383} + t_{579},
	t_{1770} = t_{940} + t_{1764},
	t_{1777} = t_{1770} + t_{1774},
	t_{1779} = t_{1777} + t_{1778},
	t_{1159} = t_{383} + t_{875},
	t_{1825} = t_{1159} + t_{1287},
	t_{1827} = t_{1729} + t_{1825},
	t_{635} = t_{383} + t_{392},
	t_{1369} = t_{627} + t_{635},
	t_{1711} = r'_{16} + t_{1369},
	t_{849} = t_{467} + t_{635},
	t_{1046} = t_{752} + t_{849},
	t_{801} = t_{389} + t_{635},
	t_{916} = t_{419} + t_{801},
	t_{496} = t_{383} + t_{398},
	t_{1410} = t_{496} + t_{798},
	t_{1916} = t_{546} + t_{1410},
	t_{1918} = t_{1908} + t_{1916},
	t_{1925} = t_{1914} + t_{1918},
	t_{1928} = t_{1925} + t_{1926},
	t_{1738} = t_{1410} + t_{1736},
	t_{1264} = t_{496} + t_{560},
	t_{1666} = t_{537} + t_{1264},
	t_{894} = t_{496} + t_{652},
	t_{563} = t_{470} + t_{496},
	t_{1064} = t_{534} + t_{563},
	t_{1781} = t_{779} + t_{1064},
	t_{1787} = t_{951} + t_{1781},
	t_{1791} = t_{1148} + t_{1787},
	t_{1794} = t_{1789} + t_{1791},
	t_{1798} = t_{775} + t_{1794},
	t_{1801} = t_{1797} + t_{1798},
	t_{1802} = t_{459} + t_{1801},
	t_{1015} = t_{563} + t_{640},
	t_{1472} = t_{518} + t_{1015},
	t_{1474} = t_{637} + t_{1472},
	t_{692} = t_{403} + t_{563},
	t_{1507} = t_{637} + t_{692},
	t_{2298} = t_{537} + t_{1507},
	t_{1811} = t_{396} + t_{1507},
	t_{991} = t_{692} + t_{926},
	t_{382} = r'_{2} + r'_{6},
	t_{1440} = r'_{167} + t_{382},
	t_{1132} = t_{382} + t_{631},
	t_{1205} = t_{629} + t_{1132},
	t_{578} = t_{382} + t_{456},
	t_{1705} = t_{578} + t_{1390},
	t_{1096} = t_{578} + t_{826},
	t_{942} = t_{533} + t_{578},
	t_{1228} = t_{489} + t_{942},
	t_{1959} = t_{1228} + t_{1958},
	t_{1961} = t_{1554} + t_{1959},
	t_{513} = t_{382} + t_{402},
	t_{1377} = t_{513} + t_{753},
	t_{1285} = t_{513} + t_{708},
	t_{1201} = t_{513} + t_{710},
	t_{1578} = t_{665} + t_{1201},
	t_{958} = t_{486} + t_{513},
	t_{58} = r'_{56} + t_{564} + t_{673} + t_{682} + t_{716} + t_{730} + t_{746} + t_{778} + t_{807} + t_{907} + t_{958} + t_{973} + t_{997} + t_{1076} + t_{1317} + t_{1457} + t_{1459} + t_{1550} + t_{1551},
	t_{2008} = t_{58} + t_{451},
	t_{2013} = t_{1580} + t_{2008},
	t_{1072} = t_{58} + t_{871},
	t_{1873} = t_{1072} + t_{1872},
	t_{1876} = t_{437} + t_{1873},
	t_{1878} = t_{1870} + t_{1876},
	t_{1884} = t_{1697} + t_{1878},
	t_{1887} = t_{1884} + t_{1886},
	t_{1890} = t_{717} + t_{1887},
	t_{1891} = t_{1889} + t_{1890},
	t_{1892} = t_{1881} + t_{1891},
	t_{1893} = t_{1707} + t_{1892},
	t_{1674} = t_{1072} + t_{1662},
	t_{1998} = t_{1674} + t_{1885},
	t_{1500} = t_{697} + t_{958},
	t_{2196} = t_{1401} + t_{1500},
	t_{2198} = t_{2193} + t_{2196},
	t_{2200} = t_{2198} + t_{2199},
	t_{2204} = t_{2200} + t_{2203},
	t_{676} = t_{482} + t_{513},
	t_{2111} = t_{676} + t_{747},
	t_{2113} = t_{611} + t_{2111},
	t_{1822} = t_{441} + t_{676},
	t_{1824} = t_{845} + t_{1822},
	t_{1826} = t_{1141} + t_{1824},
	t_{1828} = t_{1826} + t_{1827},
	t_{1543} = t_{522} + t_{676},
	t_{1327} = t_{573} + t_{676},
	t_{917} = t_{505} + t_{676},
	t_{1462} = t_{917} + t_{1079},
	t_{1165} = t_{534} + t_{917},
	t_{1590} = t_{891} + t_{1165},
	t_{381} = r'_{11} + r'_{15},
	t_{691} = t_{381} + t_{477},
	t_{1355} = t_{622} + t_{691},
	t_{1256} = t_{691} + t_{822},
	t_{1461} = t_{636} + t_{1256},
	t_{1250} = r'_{124} + t_{691},
	t_{1311} = t_{1000} + t_{1250},
	t_{1586} = t_{1311} + t_{1585},
	t_{850} = t_{607} + t_{691},
	t_{1948} = t_{850} + t_{1947},
	t_{1950} = t_{462} + t_{1948},
	t_{1952} = t_{1946} + t_{1950},
	t_{1953} = t_{1951} + t_{1952},
	t_{1955} = t_{974} + t_{1953},
	t_{1001} = t_{421} + t_{850},
	t_{2030} = t_{1001} + t_{1377},
	t_{2036} = t_{1644} + t_{2030},
	t_{2037} = t_{875} + t_{2036},
	t_{2038} = t_{2035} + t_{2037},
	t_{2039} = t_{924} + t_{2038},
	t_{2040} = t_{1600} + t_{2039},
	t_{1722} = t_{806} + t_{1001},
	t_{495} = t_{381} + t_{390},
	t_{1518} = t_{495} + t_{1155},
	t_{1393} = t_{430} + t_{495},
	t_{1808} = t_{629} + t_{1393},
	t_{1814} = t_{418} + t_{1808},
	t_{1818} = t_{1811} + t_{1814},
	t_{1296} = t_{495} + t_{536},
	t_{1668} = t_{686} + t_{1296},
	t_{777} = t_{433} + t_{495},
	t_{1358} = t_{624} + t_{777},
	t_{1934} = t_{692} + t_{1358},
	t_{1936} = t_{559} + t_{1934},
	t_{1754} = t_{1358} + t_{1752},
	t_{823} = t_{446} + t_{777},
	t_{1949} = t_{823} + t_{1462},
	t_{1956} = t_{590} + t_{1949},
	t_{1962} = t_{1956} + t_{1957},
	t_{1829} = t_{823} + t_{1828},
	t_{1831} = t_{569} + t_{1829},
	t_{1544} = t_{823} + t_{896},
	t_{568} = t_{410} + t_{495},
	t_{1006} = t_{490} + t_{568},
	t_{1727} = t_{1006} + t_{1703},
	t_{1253} = t_{620} + t_{1006},
	t_{834} = t_{492} + t_{568},
	t_{2393} = t_{579} + t_{834},
	t_{2402} = t_{2393} + t_{2401},
	t_{2408} = r'_{104} + t_{2402},
	t_{2415} = t_{2408} + t_{2410},
	t_{1077} = t_{629} + t_{834},
	t_{1396} = t_{466} + t_{1077},
	t_{1728} = t_{1396} + t_{1590},
	t_{380} = r'_{85} + r'_{89},
	t_{1012} = t_{380} + t_{925},
	t_{1786} = t_{1012} + t_{1785},
	t_{1792} = t_{1786} + t_{1790},
	t_{1793} = t_{1100} + t_{1792},
	t_{1795} = t_{1343} + t_{1793},
	t_{1799} = t_{1788} + t_{1795},
	t_{1800} = t_{1330} + t_{1799},
	t_{1803} = t_{1783} + t_{1800},
	t_{1804} = t_{781} + t_{1803},
	t_{1805} = t_{1796} + t_{1804},
	t_{1806} = t_{1802} + t_{1805},
	t_{690} = t_{380} + t_{405},
	t_{1657} = r'_{36} + t_{690},
	t_{1854} = t_{1657} + t_{1851},
	t_{1855} = t_{592} + t_{1854},
	t_{1858} = t_{1054} + t_{1855},
	t_{1861} = t_{1857} + t_{1858},
	t_{1864} = t_{1853} + t_{1861},
	t_{1866} = t_{1088} + t_{1864},
	t_{1867} = t_{1859} + t_{1866},
	t_{1407} = t_{586} + t_{690},
	t_{1173} = t_{690} + t_{699},
	t_{1126} = t_{690} + t_{840},
	t_{829} = t_{513} + t_{690},
	t_{1696} = t_{829} + t_{929},
	t_{1207} = t_{778} + t_{829},
	t_{1748} = t_{636} + t_{1207},
	t_{1749} = t_{517} + t_{1748},
	t_{501} = t_{380} + t_{387},
	t_{1109} = t_{501} + t_{532},
	t_{1671} = t_{507} + t_{1109},
	t_{1536} = t_{473} + t_{1109},
	t_{1932} = t_{1536} + t_{1780},
	t_{734} = t_{486} + t_{501},
	t_{1435} = t_{734} + t_{1023},
	t_{2154} = t_{1088} + t_{1435},
	t_{2367} = t_{2154} + t_{2361},
	t_{2372} = t_{1413} + t_{2367},
	t_{1142} = t_{568} + t_{734},
	t_{63} = t_{61} + t_{513} + t_{746} + t_{1142} + t_{1427} + t_{1474} + t_{1511} + t_{1567} + t_{1618},
	t_{2429} = t_{63} + t_{2428},
	t_{2431} = t_{452} + t_{2429},
	t_{2432} = t_{2430} + t_{2431},
	t_{2433} = t_{1537} + t_{2432},
	t_{23} = t_{596} + t_{600} + t_{668} + t_{741} + t_{760} + t_{778} + t_{844} + t_{916} + t_{966} + t_{1009} + t_{1116} + t_{1428} + t_{1461} + t_{2433},
	t_{1751} = t_{23} + t_{1749},
	t_{1755} = t_{697} + t_{1751},
	t_{1756} = t_{1754} + t_{1755},
	t_{1238} = t_{23} + t_{1019},
	t_{122} = t_{404} + t_{422} + t_{436} + t_{447} + t_{635} + t_{666} + t_{689} + t_{844} + t_{878} + t_{895} + t_{910} + t_{930} + t_{1182} + t_{1238} + t_{1649} + t_{1670} + t_{1730} + t_{1895},
	t_{0} = t_{608} + t_{2433},
	t_{1106} = t_{0} + t_{63},
	t_{1175} = t_{906} + t_{1106},
	t_{626} = t_{0} + t_{23},
	t_{744} = t_{608} + t_{626},
	t_{54} = t_{381} + t_{504} + t_{624} + t_{744} + t_{761} + t_{849} + t_{899} + t_{1012} + t_{1029} + t_{1228} + t_{1335} + t_{1515} + t_{1592} + t_{1631} + t_{1670},
	t_{2032} = t_{479} + t_{744},
	t_{2033} = t_{2031} + t_{2032},
	t_{2041} = t_{2033} + t_{2040},
	t_{5} = t_{1143} + t_{2041},
	t_{1594} = t_{5} + t_{946},
	t_{1691} = t_{1436} + t_{1594},
	t_{1138} = t_{400} + t_{744},
	t_{53} = t_{449} + t_{468} + t_{487} + t_{518} + t_{530} + t_{607} + t_{668} + t_{699} + t_{763} + t_{902} + t_{921} + t_{959} + t_{1019} + t_{1138} + t_{1146} + t_{1258} + t_{1473} + t_{1631},
	t_{1208} = t_{53} + t_{686},
	t_{1665} = t_{1035} + t_{1208},
	t_{8} = t_{390} + t_{408} + t_{415} + t_{427} + t_{484} + t_{555} + t_{583} + t_{690} + t_{818} + t_{941} + t_{1138} + t_{1397} + t_{1649} + t_{1686} + t_{1687},
	t_{1603} = t_{63} + t_{1138},
	t_{99} = t_{434} + t_{435} + t_{464} + t_{476} + t_{479} + t_{507} + t_{525} + t_{620} + t_{739} + t_{1393} + t_{1603} + t_{1619} + t_{1730},
	t_{1960} = t_{99} + t_{743},
	t_{1963} = t_{1960} + t_{1962},
	t_{1965} = t_{1963} + t_{1964},
	t_{1966} = t_{1961} + t_{1965},
	t_{1967} = t_{1505} + t_{1966},
	t_{55} = t_{1327} + t_{1967},
	t_{1314} = t_{55} + t_{555},
	t_{1504} = t_{1018} + t_{1314},
	t_{91} = t_{517} + t_{530} + t_{541} + t_{600} + t_{628} + t_{838} + t_{883} + t_{1012} + t_{1196} + t_{1452} + t_{1465} + t_{1504} + t_{1992},
	t_{1202} = r'_{247} + t_{55},
	t_{1823} = t_{628} + t_{1202},
	t_{1830} = t_{1531} + t_{1823},
	t_{1832} = t_{1830} + t_{1831},
	t_{1833} = t_{1473} + t_{1832},
	t_{2310} = t_{394} + t_{1967},
	t_{1483} = t_{99} + t_{445},
	t_{30} = t_{53} + t_{380} + t_{426} + t_{474} + t_{652} + t_{915} + t_{963} + t_{1209} + t_{1366} + t_{1397} + t_{1483} + t_{1603} + t_{1988} + t_{2194},
	t_{1810} = t_{30} + t_{1642},
	t_{1812} = t_{1705} + t_{1810},
	t_{1813} = t_{99} + t_{1812},
	t_{1815} = t_{1238} + t_{1813},
	t_{1816} = t_{1753} + t_{1815},
	t_{1817} = t_{498} + t_{1816},
	t_{1819} = t_{896} + t_{1817},
	t_{1757} = t_{1483} + t_{1756},
	t_{24} = t_{419} + t_{719} + t_{1757},
	t_{1809} = t_{1757} + t_{1807},
	t_{1820} = t_{1809} + t_{1819},
	t_{1821} = t_{1818} + t_{1820},
	t_{113} = t_{960} + t_{1821},
	t_{32} = t_{575} + t_{662} + t_{790} + t_{840} + t_{1061} + t_{1064} + t_{1258} + t_{1696} + t_{1821} + t_{1949} + t_{2169} + t_{2187},
	t_{605} = t_{61} + t_{63},
	t_{1016} = t_{605} + t_{741},
	t_{873} = t_{578} + t_{605},
	t_{1367} = t_{873} + t_{916},
	t_{2108} = t_{446} + t_{1367},
	t_{1279} = t_{620} + t_{873},
	t_{1261} = t_{394} + t_{1142},
	t_{1896} = t_{531} + t_{1261},
	t_{1902} = t_{556} + t_{1896},
	t_{685} = t_{501} + t_{516},
	t_{1180} = t_{551} + t_{685},
	t_{1271} = r'_{30} + t_{1180},
	t_{1151} = t_{526} + t_{685},
	t_{816} = t_{631} + t_{685},
	t_{1262} = t_{816} + t_{928},
	t_{121} = t_{113} + t_{382} + t_{480} + t_{500} + t_{543} + t_{782} + t_{975} + t_{1111} + t_{1157} + t_{1262} + t_{1447} + t_{1714},
	t_{653} = t_{113} + t_{121},
	t_{1289} = t_{382} + t_{653},
	t_{2044} = t_{877} + t_{1289},
	t_{879} = t_{531} + t_{653},
	t_{92} = t_{32} + t_{456} + t_{525} + t_{526} + t_{734} + t_{766} + t_{821} + t_{879} + t_{992} + t_{1086} + t_{1302} + t_{1554} + t_{2385},
	t_{1552} = t_{92} + t_{121},
	t_{66} = t_{738} + t_{1552} + t_{2204},
	t_{554} = t_{32} + t_{92},
	t_{108} = t_{489} + t_{554} + t_{1955},
	t_{2400} = t_{108} + t_{680},
	t_{1679} = t_{108} + t_{538},
	t_{1615} = t_{554} + t_{1151},
	t_{2382} = t_{1615} + t_{1696},
	t_{2384} = t_{1004} + t_{2382},
	t_{2386} = t_{1413} + t_{2384},
	t_{2308} = t_{1615} + t_{2306},
	t_{2312} = t_{2308} + t_{2310},
	t_{2313} = t_{2309} + t_{2312},
	t_{765} = t_{546} + t_{554},
	t_{114} = t_{122} + t_{513} + t_{532} + t_{685} + t_{765} + t_{928} + t_{1339} + t_{1628} + t_{1683},
	t_{617} = t_{114} + t_{122},
	t_{2049} = t_{523} + t_{617},
	t_{2307} = t_{692} + t_{2049},
	t_{726} = t_{557} + t_{617},
	t_{2422} = t_{765} + t_{2421},
	t_{2423} = t_{553} + t_{2422},
	t_{2424} = t_{2420} + t_{2423},
	t_{1135} = t_{765} + t_{868},
	t_{1203} = t_{521} + t_{1135},
	t_{6} = t_{438} + t_{498} + t_{518} + t_{691} + t_{796} + t_{900} + t_{1126} + t_{1203} + t_{1269} + t_{1293} + t_{1328} + t_{1453} + t_{1554} + t_{1665} + t_{1679},
	t_{1476} = t_{6} + t_{1030},
	t_{72} = r'_{180} + t_{1476} + t_{1893},
	t_{1745} = t_{1203} + t_{1743},
	t_{1734} = t_{410} + t_{879},
	t_{1739} = t_{1411} + t_{1734},
	t_{1741} = t_{542} + t_{1739},
	t_{1363} = t_{879} + t_{1016},
	t_{2383} = t_{1363} + t_{1609},
	t_{2387} = t_{2383} + t_{2386},
	t_{2390} = t_{2387} + t_{2388},
	t_{2391} = t_{2389} + t_{2390},
	t_{97} = t_{390} + t_{2391},
	t_{969} = t_{97} + t_{746},
	t_{56} = t_{385} + t_{486} + t_{536} + t_{590} + t_{653} + t_{969} + t_{1046} + t_{1099} + t_{1141} + t_{1181} + t_{1279} + t_{1288} + t_{1678} + t_{1679} + t_{1686} + t_{2041},
	t_{1509} = t_{56} + t_{649},
	t_{1613} = t_{969} + t_{1106},
	t_{22} = t_{465} + t_{520} + t_{560} + t_{608} + t_{1029} + t_{1474} + t_{1613} + t_{1619},
	t_{1744} = t_{22} + t_{1741},
	t_{1154} = t_{22} + t_{97},
	t_{64} = t_{466} + t_{697} + t_{790} + t_{822} + t_{838} + t_{918} + t_{939} + t_{1083} + t_{1141} + t_{1154} + t_{1285} + t_{1312} + t_{1954},
	t_{1466} = t_{64} + t_{1254},
	t_{1448} = t_{64} + t_{1407},
	t_{1740} = t_{467} + t_{1448},
	t_{1742} = t_{1738} + t_{1740},
	t_{1746} = t_{1742} + t_{1745},
	t_{1747} = t_{1744} + t_{1746},
	t_{100} = t_{605} + t_{1747},
	t_{1442} = t_{451} + t_{1154},
	t_{2212} = t_{1442} + t_{1543},
	t_{2295} = t_{1262} + t_{1427},
	t_{2297} = t_{1017} + t_{2295},
	t_{1136} = t_{816} + t_{942},
	t_{2206} = t_{1027} + t_{1136},
	t_{2209} = t_{1535} + t_{2206},
	t_{1431} = t_{484} + t_{1136},
	t_{2418} = t_{838} + t_{1431},
	t_{2419} = t_{1509} + t_{2418},
	t_{2425} = t_{2419} + t_{2424},
	t_{2426} = t_{537} + t_{2425},
	t_{74} = t_{1625} + t_{2426},
	t_{880} = t_{74} + t_{726},
	t_{2046} = t_{880} + t_{1196},
	t_{1091} = t_{554} + t_{880},
	t_{98} = t_{64} + t_{389} + t_{415} + t_{426} + t_{482} + t_{752} + t_{917} + t_{921} + t_{975} + t_{1091} + t_{1264} + t_{1327} + t_{1335} + t_{1428} + t_{1509} + t_{2389},
	t_{1931} = t_{98} + t_{1215},
	t_{1939} = t_{1931} + t_{1936},
	t_{1940} = t_{1935} + t_{1939},
	t_{1937} = t_{1091} + t_{1930},
	t_{1938} = t_{1933} + t_{1937},
	t_{1941} = t_{1938} + t_{1940},
	t_{1942} = t_{1932} + t_{1941},
	t_{1943} = t_{1579} + t_{1942},
	t_{1944} = t_{1466} + t_{1943},
	t_{89} = t_{488} + t_{1944},
	t_{1048} = t_{89} + t_{779},
	t_{833} = t_{56} + t_{89},
	t_{37} = t_{419} + t_{427} + t_{471} + t_{491} + t_{533} + t_{540} + t_{833} + t_{835} + t_{882} + t_{921} + t_{956} + t_{997} + t_{1009} + t_{1018} + t_{1453} + t_{1515} + t_{1518},
	t_{29} = t_{747} + t_{833} + t_{991} + t_{1045} + t_{1465} + t_{1510} + t_{1606} + t_{2426},
	t_{1499} = t_{29} + t_{833},
	t_{76} = t_{54} + t_{518} + t_{531} + t_{551} + t_{928} + t_{979} + t_{1064} + t_{1091} + t_{1132} + t_{1413} + t_{1499},
	t_{749} = t_{54} + t_{76},
	t_{1834} = t_{607} + t_{749},
	t_{1836} = t_{1135} + t_{1834},
	t_{1842} = t_{91} + t_{1836},
	t_{1844} = t_{1840} + t_{1842},
	t_{1845} = t_{726} + t_{1844},
	t_{1846} = t_{1841} + t_{1845},
	t_{1847} = t_{1843} + t_{1846},
	t_{1848} = t_{1504} + t_{1847},
	t_{39} = t_{1205} + t_{1848},
	t_{1589} = t_{640} + t_{749},
	t_{90} = t_{30} + t_{526} + t_{680} + t_{806} + t_{995} + t_{1431} + t_{1589} + t_{1624},
	t_{757} = t_{30} + t_{90},
	t_{2249} = t_{717} + t_{757},
	t_{106} = t_{477} + t_{629} + t_{729} + t_{736} + t_{790} + t_{1034} + t_{1254} + t_{1531} + t_{2249},
	t_{1545} = r'_{247} + t_{106},
	t_{1688} = r'_{79} + t_{1545},
	t_{12} = r'_{37} + r'_{171} + r'_{226} + t_{529} + t_{705} + t_{834} + t_{842} + t_{888} + t_{913} + t_{923} + t_{1081} + t_{1209} + t_{1215} + t_{1336} + t_{1385} + t_{1450} + t_{1566} + t_{1638} + t_{1652} + t_{1685} + t_{1688},
	t_{1593} = t_{12} + t_{537},
	t_{102} = r'_{118} + r'_{140} + r'_{157} + r'_{187} + r'_{241} + t_{411} + t_{568} + t_{907} + t_{967} + t_{1103} + t_{1139} + t_{1274} + t_{1310} + t_{1332} + t_{1395} + t_{1398} + t_{1403} + t_{1412} + t_{1492} + t_{1516} + t_{1533} + t_{1535} + t_{1581} + t_{1593} + t_{1688} + t_{1720},
	t_{1322} = t_{106} + t_{956},
	t_{1723} = r'_{20} + t_{1322},
	t_{2245} = t_{1723} + t_{2244},
	t_{2246} = t_{2241} + t_{2245},
	t_{2255} = t_{2243} + t_{2246},
	t_{2250} = t_{2248} + t_{2249},
	t_{2257} = t_{2250} + t_{2256},
	t_{2259} = t_{2257} + t_{2258},
	t_{2260} = r'_{148} + t_{2259},
	t_{2262} = t_{2255} + t_{2260},
	t_{2263} = t_{1233} + t_{2262},
	t_{2311} = t_{1589} + t_{2307},
	t_{2314} = t_{2311} + t_{2313},
	t_{75} = t_{494} + t_{2314},
	t_{641} = t_{55} + t_{75},
	t_{86} = r'_{26} + r'_{28} + r'_{67} + r'_{71} + r'_{91} + r'_{116} + r'_{120} + r'_{132} + r'_{183} + r'_{229} + t_{102} + t_{434} + t_{641} + t_{668} + t_{768} + t_{791} + t_{806} + t_{1007} + t_{1043} + t_{1096} + t_{1111} + t_{1260} + t_{1282} + t_{1547} + t_{1575} + t_{1611} + t_{1620},
	t_{1113} = t_{557} + t_{641},
	t_{31} = t_{91} + t_{700} + t_{734} + t_{1113} + t_{1511} + t_{1624} + t_{1628} + t_{1666},
	t_{661} = t_{31} + t_{91},
	t_{107} = t_{471} + t_{497} + t_{504} + t_{565} + t_{610} + t_{661} + t_{746} + t_{883} + t_{1174} + t_{1241} + t_{1692},
	t_{2369} = r'_{37} + t_{107},
	t_{2373} = t_{2369} + t_{2370},
	t_{2374} = t_{2372} + t_{2373},
	t_{1894} = t_{107} + t_{540},
	t_{1245} = t_{107} + t_{981},
	t_{2006} = t_{1245} + t_{2005},
	t_{2012} = r'_{105} + t_{2006},
	t_{2014} = t_{2012} + t_{2013},
	t_{2015} = t_{2011} + t_{2014},
	t_{78} = t_{1728} + t_{1885} + t_{2015},
	t_{1160} = r'_{42} + t_{78},
	t_{1616} = t_{1160} + t_{1372},
	t_{1706} = r'_{191} + t_{1616},
	t_{1306} = r'_{34} + t_{1245},
	t_{80} = r'_{30} + r'_{139} + t_{407} + t_{432} + t_{442} + t_{464} + t_{607} + t_{663} + t_{724} + t_{742} + t_{808} + t_{809} + t_{923} + t_{938} + t_{1096} + t_{1171} + t_{1200} + t_{1306} + t_{1321} + t_{1327} + t_{1523} + t_{1607} + t_{1653} + t_{1720},
	t_{2186} = t_{1306} + t_{2184},
	t_{2188} = t_{2179} + t_{2186},
	t_{898} = t_{107} + t_{551},
	t_{1364} = r'_{182} + t_{898},
	t_{1643} = t_{514} + t_{1364},
	t_{36} = t_{901} + t_{1435} + t_{1643} + t_{1706} + t_{2167},
	t_{2164} = t_{1501} + t_{1643},
	t_{2375} = t_{2164} + t_{2374},
	t_{2377} = t_{1047} + t_{2375},
	t_{750} = t_{107} + t_{473},
	t_{2185} = t_{750} + t_{2176},
	t_{2190} = t_{2185} + t_{2188},
	t_{1325} = t_{414} + t_{750},
	t_{123} = t_{99} + t_{439} + t_{446} + t_{522} + t_{640} + t_{809} + t_{930} + t_{952} + t_{1173} + t_{1174} + t_{1178} + t_{1232} + t_{1293} + t_{1325} + t_{1339} + t_{1694} + t_{2314},
	t_{1093} = t_{64} + t_{750},
	t_{2205} = t_{1093} + t_{1822},
	t_{1463} = t_{486} + t_{1093},
	t_{904} = t_{641} + t_{661},
	t_{1549} = t_{904} + t_{1463},
	t_{62} = r'_{247} + t_{649} + t_{711} + t_{725} + t_{829} + t_{904} + t_{992} + t_{1090} + t_{1632} + t_{1666} + t_{1684} + t_{1747} + t_{1841} + t_{2391},
	t_{82} = r'_{108} + r'_{121} + r'_{135} + r'_{170} + r'_{239} + t_{62} + t_{571} + t_{667} + t_{764} + t_{1019} + t_{1032} + t_{1200} + t_{1227} + t_{1433} + t_{1522} + t_{1718} + t_{1998} + t_{2015} + t_{2164},
	t_{562} = t_{62} + t_{82},
	t_{71} = t_{562} + t_{2148},
	t_{1897} = t_{562} + t_{1894},
	t_{21} = t_{62} + t_{521} + t_{584} + t_{773} + t_{801} + t_{900} + t_{939} + t_{1016} + t_{1461} + t_{1500},
	t_{1429} = t_{21} + t_{883},
	t_{1712} = t_{62} + t_{1429},
	t_{2109} = t_{1712} + t_{2108},
	t_{2110} = t_{2107} + t_{2109},
	t_{2114} = t_{2110} + t_{2112},
	t_{2115} = t_{2113} + t_{2114},
	t_{40} = t_{108} + t_{391} + t_{433} + t_{569} + t_{596} + t_{631} + t_{684} + t_{734} + t_{746} + t_{827} + t_{868} + t_{896} + t_{929} + t_{1104} + t_{1182} + t_{1557} + t_{1722} + t_{2115},
	t_{876} = r'_{20} + t_{21},
	t_{1639} = t_{841} + t_{876},
	t_{96} = t_{1639} + t_{1779},
	t_{1075} = t_{96} + t_{643},
	t_{18} = r'_{102} + r'_{119} + r'_{121} + r'_{194} + r'_{195} + r'_{228} + r'_{234} + t_{21} + t_{466} + t_{591} + t_{721} + t_{727} + t_{776} + t_{785} + t_{795} + t_{902} + t_{954} + t_{1052} + t_{1053} + t_{1075} + t_{1118} + t_{1133} + t_{1286} + t_{1480} + t_{1494} + t_{1556} + t_{1573} + t_{1634} + t_{1729},
	t_{1224} = t_{973} + t_{1075},
	t_{971} = t_{860} + t_{876},
	t_{1417} = t_{470} + t_{971},
	t_{59} = r'_{162} + t_{501} + t_{649} + t_{794} + t_{808} + t_{893} + t_{895} + t_{931} + t_{965} + t_{1224} + t_{1252} + t_{1286} + t_{1392} + t_{1417} + t_{1529},
	t_{20} = r'_{15} + r'_{98} + r'_{101} + r'_{225} + t_{59} + t_{441} + t_{520} + t_{606} + t_{657} + t_{665} + t_{699} + t_{722} + t_{889} + t_{892} + t_{899} + t_{1122} + t_{1331} + t_{1449} + t_{1471} + t_{1491} + t_{1566} + t_{1601} + t_{1647} + t_{1672} + t_{1983},
	t_{1700} = t_{59} + t_{517},
	t_{1058} = t_{59} + t_{808},
	t_{1292} = t_{391} + t_{1058},
	t_{2396} = t_{933} + t_{1292},
	t_{2407} = t_{2396} + t_{2404},
	t_{2409} = t_{1037} + t_{2407},
	t_{2411} = t_{1523} + t_{2409},
	t_{2413} = t_{2400} + t_{2411},
	t_{2414} = t_{2405} + t_{2413},
	t_{2416} = t_{2414} + t_{2415},
	t_{2417} = t_{2403} + t_{2416},
	t_{10} = t_{650} + t_{772} + t_{2417},
	t_{2221} = r'_{32} + t_{10},
	t_{2223} = t_{1587} + t_{2221},
	t_{2232} = t_{1591} + t_{2223},
	t_{1185} = t_{10} + t_{627},
	t_{1708} = t_{564} + t_{1185},
	t_{2017} = t_{632} + t_{1708},
	t_{2025} = t_{1565} + t_{2017},
	t_{2026} = t_{2020} + t_{2025},
	t_{2029} = t_{2026} + t_{2028},
	t_{77} = t_{1539} + t_{2029},
	t_{2057} = t_{565} + t_{2029},
	t_{2059} = t_{891} + t_{2057},
	t_{2064} = t_{2056} + t_{2059},
	t_{2068} = t_{2064} + t_{2067},
	t_{2069} = t_{1108} + t_{2068},
	t_{2070} = t_{2060} + t_{2069},
	t_{103} = t_{633} + t_{2070},
	t_{87} = r'_{1} + r'_{29} + r'_{112} + r'_{131} + t_{536} + t_{587} + t_{1150} + t_{1152} + t_{1159} + t_{1179} + t_{1229} + t_{1289} + t_{1351} + t_{1423} + t_{1432} + t_{1564} + t_{1568} + t_{1604} + t_{1611} + t_{1656} + t_{2070},
	t_{1375} = t_{381} + t_{1292},
	t_{120} = r'_{0} + r'_{10} + r'_{74} + r'_{109} + r'_{112} + r'_{149} + r'_{212} + r'_{236} + t_{86} + t_{543} + t_{577} + t_{614} + t_{625} + t_{688} + t_{807} + t_{846} + t_{1010} + t_{1045} + t_{1049} + t_{1126} + t_{1252} + t_{1308} + t_{1348} + t_{1375} + t_{1434} + t_{1454} + t_{1479} + t_{1491} + t_{1779},
	t_{789} = t_{86} + t_{120},
	t_{1710} = t_{658} + t_{789},
	t_{2315} = t_{1355} + t_{1710},
	t_{2317} = t_{1099} + t_{2315},
	t_{2319} = t_{976} + t_{2317},
	t_{2330} = t_{547} + t_{2319},
	t_{2333} = t_{2328} + t_{2330},
	t_{2334} = t_{2332} + t_{2333},
	t_{85} = r'_{4} + r'_{44} + r'_{252} + t_{391} + t_{483} + t_{655} + t_{788} + t_{797} + t_{821} + t_{836} + t_{892} + t_{1144} + t_{1190} + t_{1257} + t_{1381} + t_{1385} + t_{1421} + t_{1438} + t_{1440} + t_{1477} + t_{1496} + t_{1532} + t_{1544} + t_{1583} + t_{1659} + t_{1693} + t_{1780} + t_{1806} + t_{2334} + t_{2412},
	t_{101} = r'_{31} + r'_{68} + r'_{128} + r'_{210} + t_{85} + t_{380} + t_{556} + t_{591} + t_{593} + t_{598} + t_{709} + t_{742} + t_{749} + t_{757} + t_{920} + t_{1046} + t_{1186} + t_{1223} + t_{1235} + t_{1275} + t_{1359} + t_{1402} + t_{1447} + t_{1512} + t_{1525} + t_{1571} + t_{1621} + t_{1690},
	t_{1163} = t_{651} + t_{789},
	t_{1451} = r'_{190} + t_{1163},
	t_{67} = r'_{205} + t_{1451} + t_{1626} + t_{2190},
	t_{104} = r'_{38} + r'_{117} + r'_{133} + r'_{158} + t_{67} + t_{536} + t_{557} + t_{585} + t_{646} + t_{696} + t_{886} + t_{915} + t_{1029} + t_{1100} + t_{1251} + t_{1395} + t_{1464} + t_{1578} + t_{1586} + t_{1659} + t_{1708} + t_{1710} + t_{1986} + t_{1992},
	t_{1541} = t_{104} + t_{1059},
	t_{1651} = r'_{148} + t_{1541},
	t_{2214} = r'_{125} + t_{1651},
	t_{2229} = t_{2214} + t_{2227},
	t_{2230} = t_{1657} + t_{2229},
	t_{2231} = t_{2219} + t_{2230},
	t_{2235} = t_{2231} + t_{2234},
	t_{2236} = t_{2232} + t_{2235},
	t_{2237} = t_{2225} + t_{2236},
	t_{2238} = t_{2222} + t_{2237},
	t_{2} = t_{92} + t_{1635} + t_{2238},
	t_{1542} = t_{62} + t_{386},
	t_{2267} = t_{555} + t_{1542},
	t_{2277} = t_{1732} + t_{2267},
	t_{2285} = t_{2277} + t_{2283},
	t_{2288} = t_{1725} + t_{2285},
	t_{2289} = t_{743} + t_{2288},
	t_{1074} = t_{62} + t_{100},
	t_{95} = r'_{146} + r'_{152} + r'_{196} + r'_{206} + t_{472} + t_{605} + t_{654} + t_{683} + t_{727} + t_{920} + t_{967} + t_{1022} + t_{1026} + t_{1036} + t_{1074} + t_{1092} + t_{1323} + t_{1346} + t_{1416} + t_{1430} + t_{1457} + t_{1613} + t_{1655} + t_{1698} + t_{1724} + t_{2082} + t_{2261},
	t_{1527} = t_{95} + t_{1175},
	t_{69} = t_{1527} + t_{2263},
	t_{1856} = t_{1074} + t_{1259},
	t_{1868} = t_{1856} + t_{1865},
	t_{1869} = t_{1867} + t_{1868},
	t_{17} = t_{1700} + t_{1869},
	t_{3} = r'_{115} + r'_{128} + r'_{144} + r'_{146} + r'_{150} + r'_{173} + r'_{204} + r'_{211} + t_{491} + t_{504} + t_{745} + t_{759} + t_{786} + t_{827} + t_{894} + t_{972} + t_{1044} + t_{1048} + t_{1053} + t_{1085} + t_{1193} + t_{1261} + t_{1277} + t_{1279} + t_{1313} + t_{1409} + t_{1574} + t_{1577} + t_{1584} + t_{1599} + t_{1617} + t_{1682} + t_{1869},
	t_{996} = r'_{212} + t_{62},
	t_{57} = r'_{48} + r'_{113} + r'_{232} + t_{386} + t_{545} + t_{602} + t_{758} + t_{865} + t_{907} + t_{971} + t_{988} + t_{996} + t_{1133} + t_{1158} + t_{1223} + t_{1224} + t_{1227} + t_{1375} + t_{1401} + t_{1468} + t_{1498} + t_{1715},
	t_{1633} = t_{667} + t_{996},
	t_{1645} = t_{57} + t_{1633},
	t_{119} = r'_{3} + r'_{76} + r'_{79} + r'_{91} + r'_{134} + r'_{173} + r'_{208} + r'_{248} + t_{555} + t_{739} + t_{935} + t_{1011} + t_{1077} + t_{1148} + t_{1152} + t_{1235} + t_{1324} + t_{1338} + t_{1384} + t_{1478} + t_{1602} + t_{1645} + t_{1655} + t_{1667} + t_{1697} + t_{1713} + t_{2217} + t_{2238},
	t_{19} = r'_{9} + r'_{253} + t_{558} + t_{825} + t_{936} + t_{1313} + t_{1355} + t_{1366} + t_{1420} + t_{1475} + t_{1479} + t_{1495} + t_{1565} + t_{1618} + t_{1645} + t_{1656} + t_{1658} + t_{1672} + t_{1693},
	t_{885} = t_{19} + t_{57},
	t_{93} = r'_{6} + r'_{45} + r'_{174} + t_{485} + t_{488} + t_{560} + t_{857} + t_{885} + t_{906} + t_{945} + t_{1014} + t_{1022} + t_{1117} + t_{1456} + t_{1471} + t_{1532} + t_{1542} + t_{1609} + t_{1617} + t_{1641} + t_{1677} + t_{1719},
	t_{1391} = r'_{1} + t_{885},
	t_{1623} = t_{1066} + t_{1391},
	t_{94} = r'_{52} + r'_{59} + r'_{138} + r'_{183} + r'_{190} + r'_{194} + r'_{241} + r'_{250} + t_{383} + t_{401} + t_{408} + t_{453} + t_{502} + t_{598} + t_{693} + t_{817} + t_{896} + t_{948} + t_{951} + t_{964} + t_{1010} + t_{1016} + t_{1150} + t_{1168} + t_{1309} + t_{1417} + t_{1584} + t_{1623} + t_{1640} + t_{1702},
	t_{2282} = t_{1033} + t_{1623},
	t_{2287} = t_{2282} + t_{2286},
	t_{2290} = t_{2287} + t_{2289},
	t_{2292} = t_{2284} + t_{2290},
	t_{2293} = t_{2276} + t_{2292},
	t_{2294} = t_{2291} + t_{2293},
	t_{26} = r'_{32} + r'_{181} + r'_{233} + t_{392} + t_{428} + t_{449} + t_{512} + t_{549} + t_{560} + t_{657} + t_{670} + t_{786} + t_{825} + t_{874} + t_{974} + t_{1009} + t_{1101} + t_{1173} + t_{1231} + t_{1272} + t_{1371} + t_{1403} + t_{1404} + t_{1493} + t_{1540} + t_{1614} + t_{1691} + t_{2270} + t_{2294},
	t_{112} = t_{26} + t_{2334},
	t_{1353} = t_{976} + t_{1113},
	t_{88} = r'_{96} + r'_{181} + t_{453} + t_{671} + t_{681} + t_{1177} + t_{1353} + t_{1452} + t_{1469} + t_{1489} + t_{1503} + t_{1553} + t_{1604} + t_{1620} + t_{1630} + t_{1651} + t_{1705},
	t_{110} = r'_{58} + r'_{70} + r'_{85} + r'_{164} + r'_{253} + t_{88} + t_{519} + t_{520} + t_{773} + t_{812} + t_{878} + t_{968} + t_{1060} + t_{1070} + t_{1172} + t_{1377} + t_{1405} + t_{1481} + t_{1502} + t_{1704} + t_{1724} + t_{1725} + t_{1954} + t_{1955} + t_{2189} + t_{2190} + t_{2412} + t_{2417},
	t_{25} = r'_{101} + r'_{106} + r'_{187} + r'_{231} + t_{382} + t_{506} + t_{597} + t_{689} + t_{693} + t_{777} + t_{789} + t_{898} + t_{925} + t_{987} + t_{1005} + t_{1020} + t_{1082} + t_{1190} + t_{1233} + t_{1288} + t_{1349} + t_{1353} + t_{1393} + t_{1394} + t_{1478} + t_{1563} + t_{1574} + t_{1575} + t_{1636} + t_{2148} + t_{2208} + t_{2270},
	t_{1439} = t_{540} + t_{749},
	t_{2042} = t_{422} + t_{1439},
	t_{115} = t_{123} + t_{413} + t_{435} + t_{661} + t_{782} + t_{822} + t_{917} + t_{1074} + t_{1363} + t_{1367} + t_{1536} + t_{1684} + t_{2042} + t_{2049},
	t_{2045} = t_{2042} + t_{2044},
	t_{2047} = t_{2045} + t_{2046},
	t_{2050} = t_{534} + t_{2047},
	t_{1319} = t_{29} + t_{878},
	t_{2118} = t_{1692} + t_{1944},
	t_{2119} = t_{2117} + t_{2118},
	t_{2120} = t_{1237} + t_{2119},
	t_{2121} = t_{1694} + t_{2120},
	t_{105} = t_{1319} + t_{2121},
	t_{1025} = t_{105} + t_{562},
	t_{16} = t_{385} + t_{394} + t_{470} + t_{531} + t_{840} + t_{959} + t_{1025} + t_{1157} + t_{1174} + t_{1537} + t_{1650},
	t_{989} = t_{16} + t_{496},
	t_{1217} = t_{544} + t_{989},
	t_{1985} = t_{1217} + t_{1984},
	t_{1991} = t_{1985} + t_{1987},
	t_{1993} = t_{1990} + t_{1991},
	t_{83} = t_{451} + t_{493} + t_{512} + t_{527} + t_{989} + t_{1253} + t_{1287} + t_{1398} + t_{1567} + t_{1848} + t_{1988} + t_{1993},
	t_{1899} = t_{1099} + t_{1217},
	t_{1900} = t_{904} + t_{1899},
	t_{1901} = t_{1897} + t_{1900},
	t_{1903} = t_{1901} + t_{1902},
	t_{1904} = t_{507} + t_{1903},
	t_{1905} = t_{1898} + t_{1904},
	t_{14} = t_{478} + t_{1513} + t_{1905},
	t_{1408} = t_{14} + t_{629},
	t_{84} = t_{16} + t_{403} + t_{458} + t_{507} + t_{689} + t_{775} + t_{784} + t_{839} + t_{855} + t_{979} + t_{1408} + t_{1549} + t_{1717} + t_{2187},
	t_{51} = r'_{132} + r'_{172} + r'_{220} + t_{84} + t_{456} + t_{470} + t_{580} + t_{600} + t_{601} + t_{671} + t_{674} + t_{860} + t_{937} + t_{957} + t_{964} + t_{1195} + t_{1206} + t_{1274} + t_{1346} + t_{1434} + t_{1448} + t_{1456} + t_{1580} + t_{1586} + t_{1621} + t_{1631} + t_{1648} + t_{1692} + t_{1704} + t_{2154},
	t_{805} = t_{84} + t_{562},
	t_{1490} = t_{805} + t_{1156},
	t_{1166} = t_{405} + t_{805},
	t_{2299} = t_{1166} + t_{2296},
	t_{2300} = t_{561} + t_{2299},
	t_{2301} = t_{2298} + t_{2300},
	t_{2302} = t_{796} + t_{2301},
	t_{2303} = t_{1287} + t_{2302},
	t_{2304} = t_{402} + t_{2303},
	t_{2305} = t_{2297} + t_{2304},
	t_{38} = t_{487} + t_{2305},
	t_{1506} = t_{38} + t_{1490},
	t_{7} = t_{1506} + t_{1833},
	t_{68} = t_{7} + t_{584} + t_{717} + t_{870} + t_{894} + t_{896} + t_{928} + t_{1002} + t_{1201} + t_{1202} + t_{1285} + t_{1288} + t_{1337} + t_{1398} + t_{1447} + t_{1449} + t_{1510} + t_{1612} + t_{1627} + t_{2296},
	t_{2207} = t_{1166} + t_{2205},
	t_{2211} = t_{2207} + t_{2209},
	t_{2213} = t_{2210} + t_{2211},
	t_{47} = t_{29} + t_{89} + t_{105} + t_{428} + t_{438} + t_{493} + t_{501} + t_{1015} + t_{1155} + t_{1460} + t_{1505} + t_{1513} + t_{1592} + t_{1722} + t_{2208} + t_{2212} + t_{2213},
	t_{15} = t_{575} + t_{1543} + t_{1555} + t_{2213},
	t_{81} = t_{15} + t_{824} + t_{878} + t_{898} + t_{975} + t_{991} + t_{1253} + t_{1467} + t_{1717} + t_{2305},
	t_{11} = r'_{0} + r'_{86} + r'_{200} + t_{81} + t_{417} + t_{515} + t_{667} + t_{702} + t_{706} + t_{730} + t_{881} + t_{888} + t_{909} + t_{926} + t_{983} + t_{1059} + t_{1071} + t_{1101} + t_{1121} + t_{1139} + t_{1195} + t_{1271} + t_{1376} + t_{1466} + t_{1577} + t_{1727} + t_{1732},
	t_{34} = r'_{193} + t_{15} + t_{405} + t_{498} + t_{567} + t_{622} + t_{655} + t_{908} + t_{922} + t_{1039} + t_{1068} + t_{1096} + t_{1129} + t_{1151} + t_{1242} + t_{1280} + t_{1338} + t_{1422} + t_{1488} + t_{1530} + t_{1578} + t_{1658} + t_{2189},
	t_{52} = r'_{49} + r'_{98} + r'_{100} + t_{34} + t_{387} + t_{434} + t_{597} + t_{661} + t_{721} + t_{750} + t_{856} + t_{955} + t_{1047} + t_{1162} + t_{1212} + t_{1220} + t_{1273} + t_{1330} + t_{1368} + t_{1370} + t_{1405} + t_{1438} + t_{1464} + t_{1600} + t_{1637} + t_{1731} + t_{1893} + t_{1998},
	t_{1907} = t_{52} + t_{1043},
	t_{1921} = t_{916} + t_{1907},
	t_{1923} = t_{1909} + t_{1921},
	t_{1924} = t_{1920} + t_{1923},
	t_{1927} = t_{1922} + t_{1924},
	t_{1929} = t_{1927} + t_{1928},
	t_{70} = r'_{8} + r'_{21} + r'_{40} + r'_{118} + r'_{127} + r'_{172} + t_{525} + t_{675} + t_{687} + t_{752} + t_{866} + t_{977} + t_{1118} + t_{1159} + t_{1240} + t_{1301} + t_{1309} + t_{1563} + t_{1583} + t_{1599} + t_{1698} + t_{1929},
	t_{9} = t_{1716} + t_{1929},
	t_{1191} = t_{9} + t_{1153},
	t_{35} = r'_{16} + r'_{30} + r'_{41} + r'_{86} + r'_{102} + r'_{110} + t_{457} + t_{648} + t_{677} + t_{723} + t_{889} + t_{1100} + t_{1179} + t_{1191} + t_{1425} + t_{1484} + t_{1548} + t_{1622} + t_{1638} + t_{1660} + t_{1671} + t_{1695} + t_{2217},
	t_{1538} = t_{35} + t_{1378},
	t_{109} = r'_{7} + r'_{152} + r'_{159} + r'_{202} + t_{462} + t_{642} + t_{679} + t_{704} + t_{754} + t_{759} + t_{811} + t_{947} + t_{1083} + t_{1207} + t_{1221} + t_{1230} + t_{1240} + t_{1265} + t_{1275} + t_{1352} + t_{1412} + t_{1423} + t_{1488} + t_{1510} + t_{1538} + t_{1576} + t_{1579} + t_{1664} + t_{1667} + t_{1685},
	t_{985} = r'_{69} + t_{35},
	t_{44} = r'_{24} + r'_{27} + r'_{72} + r'_{98} + r'_{117} + r'_{176} + r'_{190} + t_{109} + t_{422} + t_{738} + t_{849} + t_{894} + t_{908} + t_{937} + t_{985} + t_{1032} + t_{1033} + t_{1041} + t_{1128} + t_{1137} + t_{1152} + t_{1278} + t_{1342} + t_{1411} + t_{1430} + t_{1544} + t_{1556} + t_{1581} + t_{1636} + t_{1727},
	t_{2362} = t_{865} + t_{985},
	t_{2365} = t_{2362} + t_{2364},
	t_{2366} = t_{627} + t_{2365},
	t_{2376} = t_{2167} + t_{2366},
	t_{2378} = t_{2376} + t_{2377},
	t_{2380} = t_{2378} + t_{2379},
	t_{2381} = t_{2371} + t_{2380},
	t_{79} = t_{1057} + t_{2381},
	t_{1291} = t_{79} + t_{1210},
	t_{1709} = t_{402} + t_{1291},
	t_{117} = r'_{222} + t_{437} + t_{483} + t_{571} + t_{644} + t_{684} + t_{726} + t_{759} + t_{761} + t_{819} + t_{877} + t_{970} + t_{1023} + t_{1145} + t_{1162} + t_{1178} + t_{1250} + t_{1268} + t_{1305} + t_{1424} + t_{1494} + t_{1553} + t_{1675} + t_{1680} + t_{1699} + t_{1709} + t_{1711},
	t_{43} = r'_{47} + r'_{90} + r'_{105} + r'_{123} + r'_{188} + r'_{253} + t_{440} + t_{453} + t_{499} + t_{533} + t_{623} + t_{901} + t_{1062} + t_{1115} + t_{1174} + t_{1225} + t_{1263} + t_{1336} + t_{1425} + t_{1480} + t_{1497} + t_{1538} + t_{1561} + t_{1647} + t_{1690} + t_{1709} + t_{1731} + t_{2263},
	t_{42} = r'_{69} + r'_{135} + r'_{141} + r'_{147} + r'_{228} + r'_{243} + t_{384} + t_{492} + t_{559} + t_{595} + t_{740} + t_{757} + t_{832} + t_{866} + t_{894} + t_{924} + t_{938} + t_{974} + t_{1127} + t_{1134} + t_{1189} + t_{1234} + t_{1299} + t_{1304} + t_{1349} + t_{1440} + t_{1454} + t_{1502} + t_{1673} + t_{1675} + t_{1676} + t_{2381},
	t_{2077} = t_{853} + t_{985},
	t_{2076} = t_{1191} + t_{1669},
	t_{2092} = t_{2076} + t_{2086},
	t_{2096} = t_{2090} + t_{2092},
	t_{2098} = t_{2073} + t_{2096},
	t_{2099} = t_{1283} + t_{2098},
	t_{2100} = t_{2077} + t_{2099},
	t_{2101} = t_{2097} + t_{2100},
	t_{2102} = t_{2089} + t_{2101},
	t_{41} = r'_{113} + r'_{240} + t_{403} + t_{480} + t_{487} + t_{521} + t_{626} + t_{651} + t_{944} + t_{1011} + t_{1030} + t_{1061} + t_{1129} + t_{1160} + t_{1277} + t_{1304} + t_{1325} + t_{1371} + t_{1415} + t_{1420} + t_{1482} + t_{1520} + t_{1525} + t_{1553} + t_{1665} + t_{1732} + t_{2102},
	t_{1} = r'_{111} + t_{2102},
	t_{45} = t_{427} + t_{439} + t_{541} + t_{873} + t_{925} + t_{1045} + t_{1046} + t_{1156} + t_{1408} + t_{1468} + t_{1524} + t_{1627} + t_{1634} + t_{1712},
	t_{118} = r'_{23} + r'_{68} + r'_{127} + r'_{167} + r'_{217} + r'_{229} + t_{45} + t_{94} + t_{381} + t_{400} + t_{713} + t_{811} + t_{932} + t_{1021} + t_{1037} + t_{1071} + t_{1094} + t_{1188} + t_{1232} + t_{1270} + t_{1370} + t_{1381} + t_{1384} + t_{1389} + t_{1469} + t_{1533} + t_{1546} + t_{1612} + t_{1641} + t_{1711} + t_{1721} + t_{1895} + t_{1905},
	t_{1284} = t_{118} + t_{570},
	t_{2341} = t_{1284} + t_{1588},
	t_{2343} = t_{1648} + t_{2341},
	t_{2357} = t_{2343} + t_{2356},
	t_{2358} = t_{2353} + t_{2357},
	t_{2360} = t_{2358} + t_{2359},
	t_{33} = t_{615} + t_{2360},
	t_{1294} = r'_{204} + t_{33},
	t_{1646} = t_{1294} + t_{1514},
	t_{50} = r'_{66} + r'_{82} + r'_{124} + r'_{186} + t_{37} + t_{82} + t_{390} + t_{493} + t_{588} + t_{596} + t_{710} + t_{713} + t_{885} + t_{988} + t_{1003} + t_{1049} + t_{1057} + t_{1067} + t_{1084} + t_{1095} + t_{1177} + t_{1260} + t_{1284} + t_{1368} + t_{1388} + t_{1444} + t_{1477} + t_{1548} + t_{1646},
	t_{27} = r'_{46} + r'_{85} + r'_{115} + r'_{138} + r'_{209} + t_{702} + t_{741} + t_{761} + t_{782} + t_{979} + t_{1020} + t_{1032} + t_{1073} + t_{1102} + t_{1148} + t_{1274} + t_{1314} + t_{1374} + t_{1443} + t_{1487} + t_{1562} + t_{1576} + t_{1646} + t_{1652} + t_{1677} + t_{1715} + t_{1833},
	t_{28} = r'_{5} + r'_{104} + r'_{110} + r'_{210} + r'_{244} + t_{461} + t_{464} + t_{494} + t_{565} + t_{626} + t_{748} + t_{856} + t_{921} + t_{961} + t_{962} + t_{1033} + t_{1263} + t_{1301} + t_{1308} + t_{1386} + t_{1459} + t_{1460} + t_{1496} + t_{1654} + t_{1689} + t_{1691} + t_{1702} + t_{2360},
	t_{1437} = t_{108} + t_{1025},
	t_{1701} = t_{440} + t_{1437},
	t_{13} = t_{980} + t_{1701} + t_{1993},
	t_{48} = t_{13} + t_{2115},
	t_{73} = t_{48} + t_{411} + t_{473} + t_{500} + t_{541} + t_{766} + t_{959} + t_{991} + t_{999} + t_{1079} + t_{1282} + t_{1668} + t_{1753} + t_{2212},
	t_{1726} = t_{73} + t_{500},
	t_{4} = t_{1204} + t_{1726} + t_{2294},
	t_{124} = t_{24} + t_{448} + t_{490} + t_{543} + t_{553} + t_{607} + t_{1027} + t_{1061} + t_{1285} + t_{1518} + t_{1596} + t_{1597} + t_{1687} + t_{2121},
	t_{111} = r'_{1} + r'_{199} + t_{124} + t_{410} + t_{629} + t_{652} + t_{767} + t_{829} + t_{847} + t_{944} + t_{1048} + t_{1134} + t_{1188} + t_{1206} + t_{1265} + t_{1271} + t_{1394} + t_{1433} + t_{1455} + t_{1543} + t_{1587} + t_{1593} + t_{1605} + t_{1669} + t_{1723} + t_{2152},
	t_{943} = t_{124} + t_{494},
	t_{2043} = t_{943} + t_{1499},
	t_{2048} = t_{1111} + t_{2043},
	t_{2051} = t_{2048} + t_{2050},
	t_{46} = t_{15} + t_{124} + t_{600} + t_{625} + t_{685} + t_{805} + t_{822} + t_{871} + t_{995} + t_{1208} + t_{1328} + t_{1426} + t_{1549} + t_{1753} + t_{2051},
	t_{116} = t_{501} + t_{719} + t_{757} + t_{2051},
	t_{65} = r'_{66} + t_{465} + t_{475} + t_{511} + t_{573} + t_{607} + t_{700} + t_{854} + t_{887} + t_{943} + t_{1048} + t_{1126} + t_{1668} + t_{1671} + t_{2194} + t_{2204},
	t_{1211} = r'_{215} + t_{943},
	t_{1345} = t_{116} + t_{1211},
	t_{1629} = t_{27} + t_{1345},
	t_{49} = t_{388} + t_{1629} + t_{1806},
	p_{186} = t_{119} + t_{121},
	p_{185} = t_{120},
	p_{184} = t_{121} + t_{123},
	p_{183} = t_{124},
	p_{182} = t_{117} + t_{119},
	p_{181} = t_{118} + t_{120},
	p_{180} = t_{123},
	p_{179} = t_{122} + t_{124},
	p_{178} = t_{109} + t_{113},
	p_{177} = t_{110},
	p_{176} = t_{113},
	p_{175} = t_{114},
	p_{174} = t_{109} + t_{111},
	p_{173} = t_{110} + t_{112},
	p_{172} = t_{115},
	p_{171} = t_{114} + t_{116},
	p_{170} = t_{103} + t_{105},
	p_{169} = t_{104},
	p_{168} = t_{104},
	p_{167} = t_{105} + t_{107} + t_{108},
	p_{166} = t_{108},
	p_{164} = t_{102} + t_{104},
	p_{165} = t_{101} + t_{103} + p_{164},
	p_{162} = t_{106} + t_{108},
	p_{163} = t_{107} + p_{162},
	p_{161} = t_{93} + t_{97},
	p_{160} = t_{94},
	p_{159} = t_{97},
	p_{158} = t_{98},
	p_{157} = t_{93} + t_{95},
	p_{156} = t_{94} + t_{96},
	p_{155} = t_{99},
	p_{154} = t_{98} + t_{100},
	p_{153} = t_{85} + t_{89},
	p_{152} = t_{86},
	p_{151} = t_{86},
	p_{150} = t_{89} + t_{90},
	p_{149} = t_{90},
	p_{147} = t_{86} + t_{88},
	p_{148} = t_{85} + t_{87} + p_{147},
	p_{145} = t_{90} + t_{92},
	p_{146} = t_{91} + p_{145},
	p_{144} = t_{79} + t_{81},
	p_{143} = t_{80},
	p_{142} = t_{80},
	p_{141} = t_{81} + t_{83} + t_{84},
	p_{140} = t_{84},
	p_{138} = t_{78} + t_{80},
	p_{139} = t_{77} + t_{79} + p_{138},
	p_{136} = t_{82} + t_{84},
	p_{137} = t_{83} + p_{136},
	p_{135} = t_{69} + t_{73},
	p_{134} = t_{70},
	p_{133} = t_{70},
	p_{132} = t_{73} + t_{74},
	p_{131} = t_{74},
	p_{129} = t_{70} + t_{72},
	p_{130} = t_{69} + t_{71} + p_{129},
	p_{127} = t_{74} + t_{76},
	p_{128} = t_{75} + p_{127},
	p_{126} = t_{65},
	p_{125} = t_{68},
	p_{124} = t_{66},
	p_{123} = t_{66},
	p_{122} = t_{67},
	p_{121} = t_{57} + t_{61},
	p_{120} = t_{58},
	p_{119} = t_{58},
	p_{118} = t_{61},
	p_{117} = t_{61} + t_{62},
	p_{116} = t_{62},
	p_{115} = t_{57} + t_{59},
	p_{113} = t_{58} + t_{60},
	p_{114} = p_{113} + p_{115},
	p_{112} = t_{63},
	p_{110} = t_{62} + t_{64},
	p_{111} = t_{63} + p_{110},
	p_{109} = t_{49} + t_{53},
	p_{108} = t_{50},
	p_{107} = t_{50},
	p_{106} = t_{53},
	p_{105} = t_{53} + t_{54},
	p_{104} = t_{54},
	p_{103} = t_{49} + t_{51},
	p_{101} = t_{50} + t_{52},
	p_{102} = p_{101} + p_{103},
	p_{100} = t_{55},
	p_{98} = t_{54} + t_{56},
	p_{99} = t_{55} + p_{98},
	p_{97} = t_{43} + t_{45},
	p_{96} = t_{44},
	p_{95} = t_{44},
	p_{94} = t_{45} + t_{47},
	p_{93} = t_{48} + p_{94},
	p_{92} = t_{48},
	p_{91} = t_{41} + t_{43},
	p_{89} = t_{42} + t_{44},
	p_{90} = p_{89} + p_{91},
	p_{88} = t_{47},
	p_{86} = t_{46} + t_{48},
	p_{87} = t_{47} + p_{86},
	p_{85} = t_{33} + t_{37},
	p_{84} = t_{34},
	p_{83} = t_{34},
	p_{82} = t_{37},
	p_{81} = t_{37} + t_{38},
	p_{80} = t_{38},
	p_{79} = t_{33} + t_{35},
	p_{77} = t_{34} + t_{36},
	p_{78} = p_{77} + p_{79},
	p_{76} = t_{39},
	p_{74} = t_{38} + t_{40},
	p_{75} = t_{39} + p_{74},
	p_{73} = t_{25} + t_{29},
	p_{72} = t_{26},
	p_{71} = t_{26},
	p_{70} = t_{27} + t_{29},
	p_{69} = t_{28},
	p_{68} = t_{28},
	p_{67} = t_{29} + t_{31} + t_{32},
	p_{66} = t_{32},
	p_{65} = t_{29},
	p_{64} = t_{29} + t_{30},
	p_{63} = t_{30},
	p_{62} = t_{25} + t_{27},
	p_{60} = t_{26} + t_{28},
	p_{61} = p_{60} + p_{62},
	p_{59} = t_{31},
	p_{57} = t_{30} + t_{32},
	p_{58} = t_{31} + p_{57},
	p_{56} = t_{17} + t_{21},
	p_{55} = t_{18},
	p_{54} = t_{19} + t_{21},
	p_{53} = t_{20},
	p_{52} = t_{20},
	p_{51} = t_{21} + t_{23},
	p_{50} = t_{24} + p_{51},
	p_{49} = t_{24},
	p_{48} = t_{21},
	p_{47} = t_{22},
	p_{46} = t_{17} + t_{19},
	p_{44} = t_{18} + t_{20},
	p_{45} = p_{44} + p_{46},
	p_{43} = t_{23},
	p_{41} = t_{22} + t_{24},
	p_{42} = t_{23} + p_{41},
	p_{40} = t_{9} + t_{13},
	p_{39} = t_{10},
	p_{38} = t_{10},
	p_{37} = t_{11} + t_{13},
	p_{36} = t_{12},
	p_{35} = t_{12},
	p_{34} = t_{13} + t_{15},
	p_{33} = t_{16} + p_{34},
	p_{32} = t_{16},
	p_{31} = t_{13},
	p_{30} = t_{13} + t_{14},
	p_{29} = t_{14},
	p_{28} = t_{9} + t_{11},
	p_{26} = t_{10} + t_{12},
	p_{27} = p_{26} + p_{28},
	p_{25} = t_{15},
	p_{23} = t_{14} + t_{16},
	p_{24} = t_{15} + p_{23},
	p_{22} = t_{3} + t_{7},
	p_{21} = t_{4} + t_{8},
	p_{20} = t_{1} + t_{5},
	p_{19} = t_{2},
	p_{18} = t_{2},
	p_{17} = t_{3} + t_{5},
	p_{16} = t_{4},
	p_{15} = t_{4},
	p_{14} = t_{5} + t_{7},
	p_{13} = t_{8} + p_{14},
	p_{12} = t_{8},
	p_{11} = t_{5},
	p_{10} = t_{5} + t_{6},
	p_{9} = t_{6},
	p_{8} = t_{7},
	p_{7} = t_{6} + t_{8},
	p_{2} = t_{7} + p_{7},
	p_{1} = p_{7},
	p_{6} = t_{1} + t_{3},
	p_{4} = t_{2} + t_{4},
	p_{5} = p_{4} + p_{6},
	p_{3} = t_{7},
	p_{0} = t_{0}.$

	Pointwise multiplication (149 multiplications): 
	$\boldsymbol{g} = \boldsymbol{p} \cdot \boldsymbol{c}$, where
	$\boldsymbol{c} = (1$, $\alpha^{5}$, $\alpha^{143}$, $\alpha^{199}$, $\alpha^{204}$, $\alpha^{136}$, $\alpha^{68}$, $\alpha^{204}$, $\alpha^{68}$, $\alpha^{38}$, $\alpha^{222}$, $\alpha^{111}$, $\alpha^{137}$, $\alpha^{183}$, $\alpha^{219}$, $\alpha^{170}$, $1$, $1$, $\alpha^{170}$, $1$, $1$, $\alpha^{170}$, $1$, $\alpha^{5}$, $\alpha^{143}$, $\alpha^{199}$, $\alpha^{204}$, $\alpha^{136}$, $\alpha^{68}$, $\alpha^{38}$, $\alpha^{222}$, $\alpha^{111}$, $\alpha^{137}$, $\alpha^{183}$, $\alpha^{219}$, $\alpha^{170}$, $1$, $1$, $\alpha^{170}$, $1$, $1$, $\alpha^{5}$, $\alpha^{143}$, $\alpha^{199}$, $\alpha^{204}$, $\alpha^{136}$, $\alpha^{68}$, $\alpha^{38}$, $\alpha^{1}$, $\alpha^{11}$, $\alpha^{137}$, $\alpha^{183}$, $\alpha^{219}$, $\alpha^{170}$, $1$, $1$, $\alpha^{170}$, $1$, $\alpha^{5}$, $\alpha^{143}$, $\alpha^{199}$, $\alpha^{204}$, $\alpha^{136}$, $\alpha^{68}$, $\alpha^{38}$, $\alpha^{222}$, $\alpha^{111}$, $\alpha^{137}$, $\alpha^{183}$, $\alpha^{170}$, $1$, $1$, $\alpha^{170}$, $1$, $1$, $\alpha^{5}$, $\alpha^{143}$, $\alpha^{199}$, $\alpha^{204}$, $\alpha^{136}$, $\alpha^{68}$, $\alpha^{38}$, $\alpha^{222}$, $\alpha^{111}$, $\alpha^{170}$, $1$, $1$, $\alpha^{5}$, $\alpha^{143}$, $\alpha^{199}$, $\alpha^{204}$, $\alpha^{136}$, $\alpha^{68}$, $\alpha^{137}$, $\alpha^{183}$, $\alpha^{219}$, $\alpha^{17}$, $1$, $1$, $1$, $\alpha^{5}$, $\alpha^{143}$, $\alpha^{199}$, $\alpha^{204}$, $\alpha^{136}$, $\alpha^{68}$, $\alpha^{38}$, $\alpha^{222}$, $\alpha^{111}$, $\alpha^{170}$, $1$, $1$, $\alpha^{5}$, $\alpha^{143}$, $\alpha^{199}$, $\alpha^{204}$, $\alpha^{136}$, $\alpha^{68}$, $\alpha^{38}$, $\alpha^{222}$, $\alpha^{111}$, $\alpha^{17}$, $1$, $1$, $1$, $\alpha^{153}$, $\alpha^{170}$, $1$, $\alpha^{68}$, $1$, $\alpha^{5}$, $\alpha^{143}$, $\alpha^{204}$, $\alpha^{136}$, $\alpha^{38}$, $\alpha^{222}$, $\alpha^{170}$, $1$, $1$, $\alpha^{5}$, $\alpha^{143}$, $\alpha^{204}$, $\alpha^{136}$, $\alpha^{137}$, $\alpha^{183}$, $\alpha^{170}$, $1$, $1$, $\alpha^{5}$, $\alpha^{143}$, $\alpha^{204}$, $\alpha^{136}$, $\alpha^{38}$, $\alpha^{222}$, $\alpha^{170}$, $1$, $1$, $\alpha^{5}$, $\alpha^{199}$, $\alpha^{204}$, $\alpha^{68}$, $\alpha^{38}$, $\alpha^{111}$, $\alpha^{170}$, $1$, $\alpha^{5}$, $\alpha^{143}$, $\alpha^{204}$, $\alpha^{136}$, $\alpha^{137}$, $\alpha^{183}$, $\alpha^{170}$, $1$, $1$, $\alpha^{5}$, $\alpha^{199}$, $\alpha^{204}$, $\alpha^{68}$, $\alpha^{38}$, $\alpha^{111}$, $\alpha^{170}$, $1$, $\alpha^{5}$, $\alpha^{199}$, $\alpha^{204}$, $\alpha^{68}$, $\alpha^{137}$, $\alpha^{219}$, $\alpha^{170}$, $1)$.

Post-additions (177 additions): $\boldsymbol{S}' = \boldsymbol{P}^T \boldsymbol{g}$. Note that in the following sequence we use $\boldsymbol{S}$ directly to avoid extra permutation.

$	t_{273} = g_{18} + g_{20},
	t_{269} = g_{110} + g_{116},
	t_{270} = g_{113} + t_{269},
	t_{271} = g_{119} + t_{270},
	t_{272} = g_{121} + t_{271},
	S_{30} = g_{111} + g_{114} + g_{117} + g_{120} + t_{272},
	t_{266} = g_{52} + g_{54},
	t_{267} = g_{49} + t_{266},
	t_{264} = g_{92} + g_{95},
	t_{262} = g_{89} + g_{97},
	t_{263} = g_{86} + t_{262},
	t_{265} = t_{263} + t_{264},
	S_{22} = g_{87} + g_{90} + g_{93} + g_{96} + t_{265},
	t_{260} = g_{32} + g_{37},
	t_{256} = g_{77} + g_{80},
	t_{255} = g_{74} + g_{83},
	t_{257} = g_{85} + t_{255},
	t_{258} = t_{256} + t_{257},
	S_{17} = g_{75} + g_{78} + g_{81} + g_{84} + t_{258},
	t_{248} = g_{98} + g_{104},
	t_{249} = g_{107} + t_{248},
	t_{250} = g_{109} + t_{249},
	t_{251} = g_{101} + t_{250},
	S_{26} = g_{99} + g_{102} + g_{105} + g_{108} + t_{251},
	t_{246} = g_{63} + g_{73},
	t_{247} = g_{71} + t_{246},
	t_{244} = g_{2} + g_{5},
	t_{241} = g_{41} + g_{44},
	t_{268} = t_{241} + t_{267},
	S_{9} = g_{42} + g_{45} + g_{50} + g_{53} + t_{268},
	t_{240} = g_{15} + g_{17},
	t_{237} = g_{57} + g_{60},
	t_{236} = g_{100} + g_{106},
	t_{242} = g_{103} + t_{236},
	S_{13} = t_{242} + t_{251},
	t_{233} = g_{1} + g_{4},
	t_{276} = g_{12} + t_{233},
	t_{277} = t_{240} + t_{276},
	S_{2} = g_{13} + g_{16} + t_{244} + t_{277},
	t_{274} = g_{9} + t_{233},
	t_{275} = t_{273} + t_{274},
	S_{8} = g_{10} + g_{19} + t_{244} + t_{275},
	t_{230} = g_{58} + g_{61},
	t_{245} = t_{230} + t_{237},
	S_{13} = g_{64} + g_{72} + t_{245} + t_{247},
	S_{11} = g_{66} + g_{67} + g_{68} + g_{69} + g_{70} + t_{245},
	t_{226} = g_{91} + g_{94},
	t_{232} = g_{88} + t_{226},
	S_{11} = t_{232} + t_{265},
	t_{225} = g_{59} + g_{62},
	t_{235} = g_{65} + t_{225},
	S_{22} = t_{235} + t_{237} + t_{247},
	t_{223} = g_{112} + g_{118},
	t_{239} = g_{115} + t_{223},
	S_{15} = t_{239} + t_{272},
	t_{222} = g_{76} + g_{82},
	t_{224} = g_{79} + t_{222},
	S_{26} = t_{224} + t_{258},
	t_{221} = g_{43} + g_{46},
	S_{18} = g_{47} + g_{48} + g_{55} + g_{56} + t_{221} + t_{241},
	t_{229} = g_{51} + t_{221},
	S_{5} = t_{229} + t_{268},
	t_{220} = g_{3} + g_{6},
	t_{234} = g_{14} + t_{220},
	S_{1} = t_{234} + t_{277},
	S_{16} = g_{7} + g_{8} + g_{21} + g_{22} + S_{1} + t_{240},
	t_{231} = g_{11} + t_{220},
	S_{4} = t_{231} + t_{275},
	t_{219} = g_{25} + g_{28},
	t_{228} = g_{27} + t_{219},
	t_{238} = g_{24} + t_{228},
	t_{227} = g_{23} + t_{219},
	t_{243} = g_{26} + t_{227},
	t_{259} = g_{35} + t_{243},
	t_{261} = t_{259} + t_{260},
	S_{6} = g_{33} + g_{36} + t_{238} + t_{261},
	S_{3} = g_{34} + t_{261},
	t_{252} = g_{29} + t_{243},
	t_{253} = g_{38} + t_{252},
	t_{254} = g_{40} + t_{253},
	S_{24} = g_{30} + g_{39} + t_{238} + t_{254},
	S_{12} = g_{31} + t_{254},
	S_{31} = g_{179} + g_{180} + g_{181} + g_{182} + g_{183} + g_{184} + g_{185} + g_{186},
	S_{29} = g_{171} + g_{172} + g_{173} + g_{174} + g_{175} + g_{176} + g_{177} + g_{178},
	S_{27} = g_{162} + g_{163} + g_{164} + g_{165} + g_{166} + g_{167} + g_{168} + g_{169} + g_{170},
	S_{25} = g_{154} + g_{155} + g_{156} + g_{157} + g_{158} + g_{159} + g_{160} + g_{161},
	S_{23} = g_{145} + g_{146} + g_{147} + g_{148} + g_{149} + g_{150} + g_{151} + g_{152} + g_{153},
	S_{21} = g_{136} + g_{137} + g_{138} + g_{139} + g_{140} + g_{141} + g_{142} + g_{143} + g_{144},
	S_{19} = g_{127} + g_{128} + g_{129} + g_{130} + g_{131} + g_{132} + g_{133} + g_{134} + g_{135},
	S_{17} = g_{122} + g_{123} + g_{124} + g_{125} + g_{126},
	S_{0} = g_{0}.$

	Overall 149 multiplications and 3970 additions over $\GF(2^8)$ are needed.

\bibliographystyle{IEEEtran}
\bibliography{IEEEabrv,rs}

\end{document}